\documentclass[useAMS,usenatbib,fleqn]{mnras} 
\pdfoutput=1
\usepackage{graphicx} 
\usepackage{times} 
\usepackage{siunitx} 
\usepackage{multirow} 
\usepackage[inline]{enumitem}
\usepackage{hyperref} 
\def\kms{\rm{ \, km \, s^{-1}}} 
 
\def\smh{\, h^{-1} \rm{ M_\odot}}

\def\mpc{\rm{ \, Mpc}} 

\def\vtru{\vec{v}^{\mathrm tru}} 
\def\vrec{\vec{v}^{\mathrm rec}} 
 
\def\vxtru{v_x^{\mathrm tru}} 
\def\vxrec{v_x^{\mathrm rec}} 
 
\def\vrtru{v_r^{\mathrm tru}} 
\def\vrrec{v_r^{\mathrm rec}}

\def\itype{{\em{I-type\ }}} 
\def\rtype{{\em{R-type\ }}} 
\def\mpch{\, h^{-1}\rm{Mpc}} 
\def\mpc{\, \rm{Mpc}}

\renewcommand{\vec}[1]{ {\bmath #1} } 

\def\aj{AJ} 
\def\araa{ARA\&A} 
\def\apj{ApJ} 
\def\apjl{ApJL} 
\def\apjs{ApJS} 
\def\ao{Appl.~Opt.} 
\def\apss{Ap\&SS} 
\def\aap{A\&A}

\def\mnras{MNRAS}

\def\prd{Phys.~Rev.~D}

\def\nat{Nature}

\def\ltsim{\lower.5ex\hbox{$\; \buildrel < \over \sim \;$}} 
\def\gtsim{\lower.5ex\hbox{$\; \buildrel > \over \sim \;$}} 
\hypersetup{ colorlinks, linkcolor=red } 

\title[Lagrangian methods for velocity reconstruction]{Performance study of Lagrangian methods: reconstruction of large scale peculiar velocities and baryonic acoustic oscillations}

\author [J. A. Keselman \& A. Nusser] { J. A. Keselman$^{1}$\thanks{E-mail: skariel@gmail.com} and A. Nusser$^{1,2}$\\
$^{1}$Physics department, Technion, Haifa 3200003, Israel\\
$^{2}$Asher Space Research Institute, Technion, Haifa 3200003, Israel }

\begin{document} 
\date{15th draft 2016 Sep 11} 
\pagerange{ 
\pageref{firstpage}-- 
\pageref{lastpage}} \pubyear{2016} 
\maketitle \label{firstpage} 
\begin{abstract}
	NoAM for ``No Action Method" is a framework for reconstructing the past orbits of observed tracers of the large scale mass density field. It seeks exact solutions of the equations of motion (EoM), satisfying initial homogeneity and the final observed particle (tracer) positions. The solutions are found iteratively reaching a specified tolerance defined as the {\rm RMS} of the distance between reconstructed and observed positions. Starting from a guess for the initial conditions, NoAM advances particles using standard N-body techniques for solving the EoM. Alternatively, the EoM can be replaced by any approximation such as Zel'dovich and second order perturbation theory (2LPT). NoAM is suitable for billions of particles and can easily handle non-regular volumes, redshift space, and other constraints. We implement NoAM to systematically compare Zel'dovich, 2LPT, and N-body dynamics over diverse configurations ranging from idealized high-res periodic simulation box to realistic galaxy mocks. Our findings are 
	\begin{enumerate*}
		\item Non-linear reconstructions with Zel'dovich, 2LPT, and full dynamics perform better than linear theory only for idealized catalogs in real space. For realistic catalogs, linear theory is the optimal choice for reconstructing velocity fields smoothed on scales $\gtsim 5\mpch$. 
		\item all non-linear back-in-time reconstructions tested here, produce comparable enhancement of the baryonic oscillation signal in the correlation function. 
	\end{enumerate*}
\end{abstract}
\begin{keywords}
	gravitation -- cosmology: theory -- dark matter -- large-scale structure of Universe 
\end{keywords}

\section{Introduction}

Observations of the low redshift ($z\ltsim 0.1$) large scale structure of the Universe include the distribution of galaxies \citep{Colless2df,Jones6df,Bilicki2013,2012ApJS..199...26H} on one hand and and their peculiar motions (deviations from a pure Hubble flow) \citep{Springob2007,CF2,Springob2014} on the other. These are two independent sets of data. In the standard paradigm of structure formation via gravitational instability, the peculiar velocity field is tightly related to the underlying mass density. An important test of the paradigm is therefore a comparison between the peculiar velocity field derived from a redshift survey to the peculiar motions inferred from catalogs of distance measurements. Linear gravitational instability theory is frequently invoked in the analyses of large scale structure observations. It offers a convenient and almost trivial relation, $\vec{v}\propto \vec{g}$ between the gravitational force field computed from a density field and the observed peculiar velocity. The theory has been frequently implemented for the analyses of the observations filtered on large physical scales ($\gtsim 10 \mpc$), where the density contrast fluctuations are still below unity. Several physical and observational effects complications are innate to  this analysis: 
\begin{enumerate*}
	\item galaxy biasing: gravitational instability allows a derivation of the velocity field from the density field of the underlying mass. However, the galaxy distribution is a biased tracer of the dark matter, and a proper modeling of the biasing of the specific type of galaxies is crucial \citep{Lavaux2007, Nusserlg14}. 
	\item redshift surveys provide the galaxy distribution in redshift space (hereafter \texttt{\bf s}{-space}). Redshifts differ from actual distances by the galaxy radial peculiar velocities. The coherence between these velocities and the unknown distribution in real (distance) space (hereafter \texttt{\bf r}-space) introduces systematic effects (redshift distortions). 
	\item inferring the density field from a redshift survey requires an accurate determination of the selection function which depends on the unknown distances of galaxies. This introduces a systematic uncertainty which can greatly affect the reconstruction of the dipole component of the velocity field at large distances \citep{K87}. 
	\item peculiar velocity catalogs are typically characterized by heterogeneous selection criteria which are hard to to match in mock catalogs designed for error analysis. 
	\item inhomogeneous Malmquist bias \citep{Lynden-Bell88} plagues any attempt to analyze these data as a function of the estimated distance coordinate. 
\end{enumerate*}
Nonetheless, an excellent match between the 2MASS Redshift Survey (2MRS) \citep{2012ApJS..199...26H} and the SFI++ \citep{Springob2007} peculiar velocity catalog have been achieved by \cite{Davis2011}. These authors worked in \texttt{\bf s}-space in order to avoid Malmquist bias and relies on realistic mock galaxy catalogs to model galaxy biasing. This result is a demonstration of the success of the gravitational instability paradigm for structure formation. 

Another powerful test which relies on the recovery of the peculiar velocity field from a redshift survey is the ``Flickering Luminosity Method" \citep{TYS,NBDL}. The method relies on the coherence between spatial variations in the distribution of galaxy luminosities, estimated from redshifts as distance proxies, with the peculiar velocity field. This method has been applied to the SDSS data \citep{SDSSDR7} and resulted on competitive constraints on the parameters \citep{Feix2015}. 

To probe smaller scales ($1-10 \mathrm Mpc$), one may need to rely on methods which incorporate non-linear gravitational effects. Extracting information from the data on all observed scales should tighten the constraints on cosmological models. In this paper we aim at a systematic comparison between different methods for the reconstruction of velocity field from a given mass distribution. These methods include linear theory, the Zel'dovich approximation and 2LPT \citep{1993MNRAS.264..375B, Bouchet95,2012JCAP...06..021R, 2014JFM...749..404Z}, and the Peebles' action method \citep{Peebles1989,Nusser2000}. 
\begin{figure}
	\includegraphics[width=231pt]{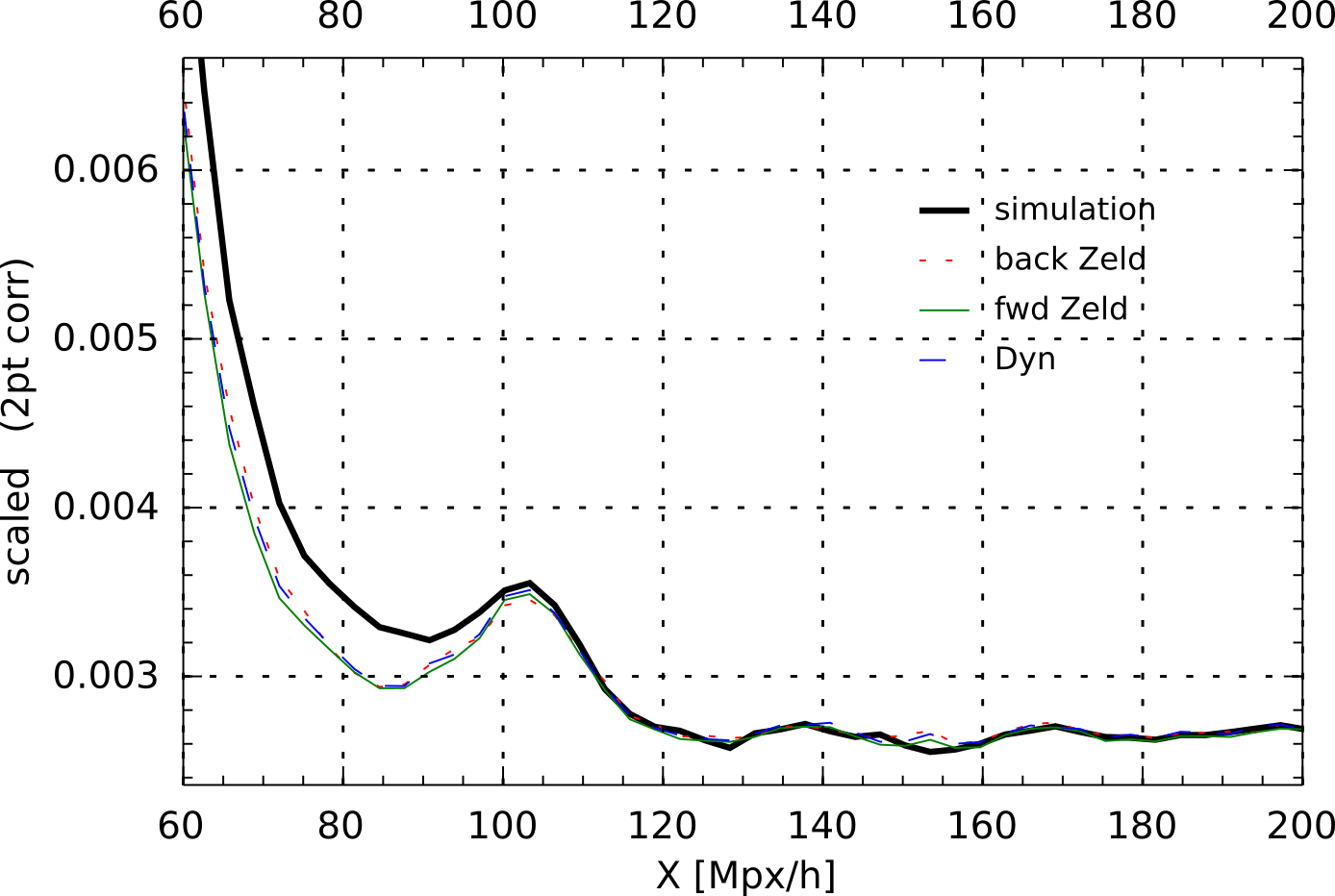} \caption{Density correlation functions computed from the \textsc{mass\_weight} \itype catalog. The thick black solid curve corresponds to the density field computed from halos in a snapshot at $z=0$. Also plotted are correlations obtained from redshift $z=0.7$. Halo distributions are reconstructed using different back-in-time methods, as indicated in the legend. Enhancement of the BAO feature at high redshift is evident. Correlations are scaled to the same power at large separations. } \label{fig:GRID_BAO} 
\end{figure}

Acoustic oscillations during the era of matter radiation coupling leave their traces in the galaxy correlation functions in terms of a peak at a $\sim 150\mathrm Mpc$ separation \citep{Sakharov66,PeeblesYu70,SunyaevZeldovich70}. This feature was detected in the 2DF \citep{ColeBAO2005} and the SDSS \citep{EisensteinBAO2005} galaxy redshift surveys. The signal is smeared by the non-linear gravitational matter displacements \citep[e.g][]{Meiksin1999}. Non-linear effects do not remove the information content associated with BAO feature, they simply cause some of this information to leak to other, higher order, statistics of the evolved density field. \cite{Eisenstein2007} argued that a reconstruction of the the density field at higher redshifts from the observed distribution will sharpen the BAO feature in the correlation function, facilitating the extraction of cosmological constraints from the observations. This requires back-in-time reconstruction methods \citep{NusserDekel92,EisensteinBAO2005,Padmanabhan2012,Kazin2014}.

The NoAM framework is presented in $\S$\ref{sec:NoAM_GOALS}, $\S$\ref{sec:EQUATIONS_OF_MOTION} and $\S$\ref{sec:THE_NOAM_METHOD}: The goal of NoAM is presented in $\S$\ref{sec:NoAM_GOALS}, the equations of motion in $\S$\ref{sec:EQUATIONS_OF_MOTION} and the iterative procedure in $\S$\ref{sec:THE_NOAM_METHOD}. The ideal and realistic mock catalogs, and the numerical methods used to solve the EoM, are presented in $\S$\ref{sec:MOCKS}. Results are presented in $\S$\ref{sec:RESULTS}.

Throughout this paper we adopt standard notation where $a(t)$ is the scale factor, $H=\dot a/a$ is the Hubble function and $\Omega(t)=\bar \rho(t)/\rho_{\mathrm c}(t)$ is the mean density of the Universe in units of the critical density, $\rho_{\mathrm c}=3H^2/8\pi G$. Further, $\vec{x}$ and $\vec{r}=a \vec{x}$ are, respectively, the comoving and physical spatial coordinates and $\vec{u}(\vec{x},t)=d\vec{x}/d t\equiv \dot \vec{x}$ is the comoving peculiar velocity and $\delta(\vec{x},t)=\rho(\vec{x},t)/\bar \rho-1$ is the density contrast.

\section{goal} \label{sec:NoAM_GOALS}

We are interested in the past history (positions as a function of time i.e. orbits) of an observed distribution of a mass tracers (galaxies). The orbits must be such that the corresponding tracer distribution approaches homogeneity as $t\rightarrow 0$. Thus, we are interested in the boundary value problem (BVP) \citep{Peebles1989,Nusserboundary} where particle positions and velocities at any time $t$ are sought under the constraints of {\em initial homogeneity} and {\em a specified distribution of matter at the present time}. Even for very accurate information on the positions and velocities at the present time, a backward solution to the equations motion will lead to strong deviations from homogeneity at early times. The most general solution to this BVP so far is provided by the Numerical Action Method (NAM) of \cite{Peebles1989}. Peebles argues that that the equations of motion render an extremum in the action for fixed final coordinates and vanishing initial peculiar velocities $a \dot \vec{x}=\vec{0}$ as $t\rightarrow 0$. The latter condition is equivalent to setting the decaying mode of linear fluctuations to zero and thus the corresponding early time density field should approach homogeneity near the initial Big Bang singularity\footnote{In general, boundary value problems have multiple solutions. To see that consider the linear oscillator $\ddot x+x=0$ subject to the boundary conditions (BC), $x(t_1=0)=0$ and $x(t_2=2\pi)=0$. There is an infinite number of solutions: $x(t)=A\sin(t)$ for any $A$. There are also BC with no solutions as is the case for $x(0)=0$ and $x(2\pi)=1$. For mixed BC where one of the conditions is $\dot x(0)=0$, the situation can be even more intriguing. For example, the BC $\dot x(0)=0 $ and $x(2\pi)=1$ are satisfied by the orbit $x(t)=\cos(t)$. Now, consider the action $S=\int_0^{2\pi}dt ({\dot x}^2-x^2)/2 $ for the orbits $x_p=\cos(t)+A \cos(w t)$. These orbits satisfy the mixed BC for $w=(2 n+1)/4$, but not the equation of motion. It is easy to see that $S=A^2\pi(w^2-1)/2$ and its extremum point (with respect to $A$) is a maximum for $w<1$ and a minimum for $w>1$.}. We will deal with the dynamical aspect to this problem. Namely, we will assume that the observations are sufficiently dense to probe the relevant scales of interest. 
\begin{figure}
	\includegraphics[width=231pt]{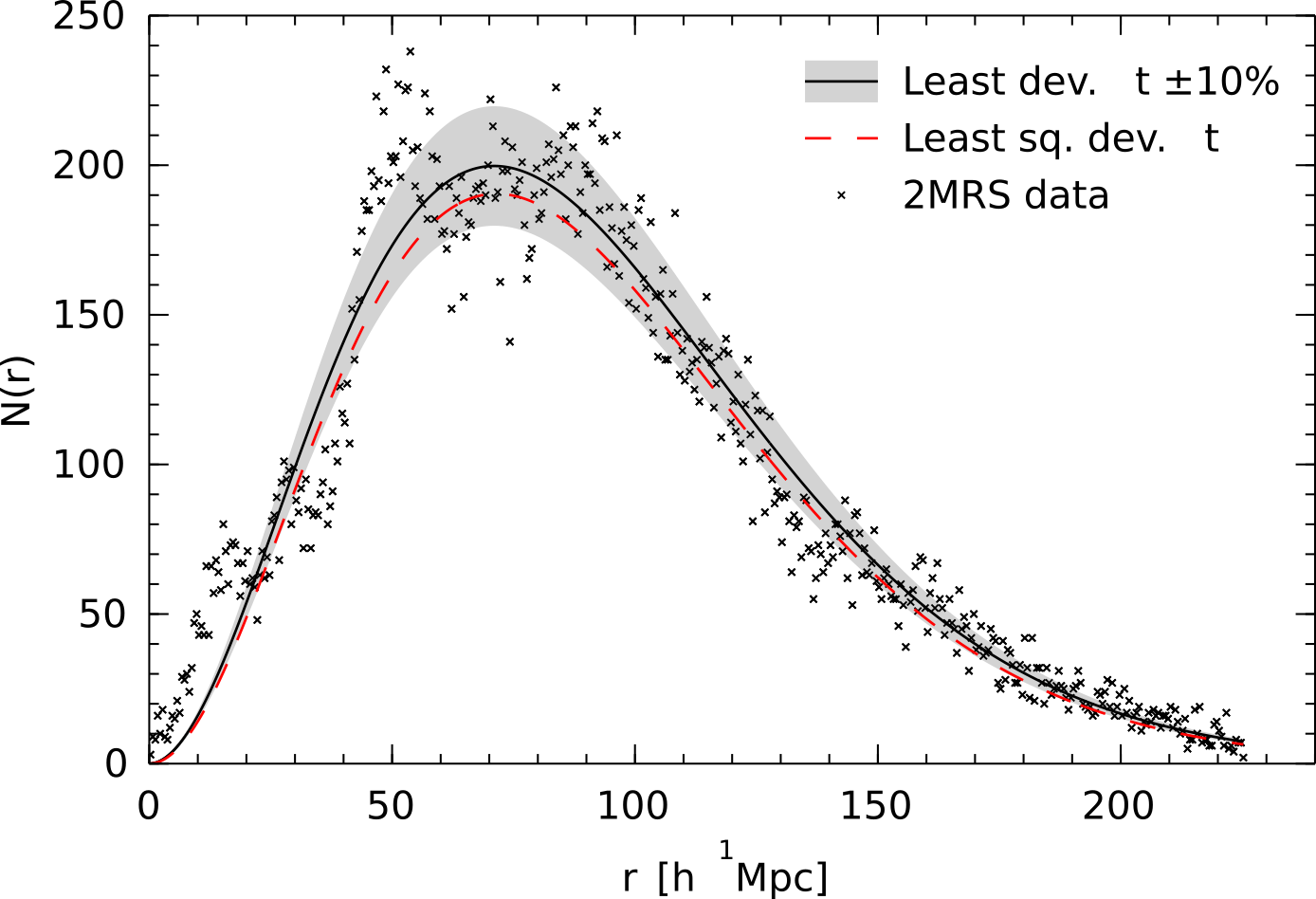} \caption{Fitting 2MRS $N(z)$ from $20$ to $225\mpch$ in bins of $0.5\mpch$. The solid curve is obtained by minimizing the absolute value of the Poisson deviations while the dashed corresponds to a least squares minimization. The grey area marks $\pm 10\%$ deviation in absolute value fit.} \label{fig:2mrs_nz} 
\end{figure}
\begin{figure}
	\includegraphics[width=231pt]{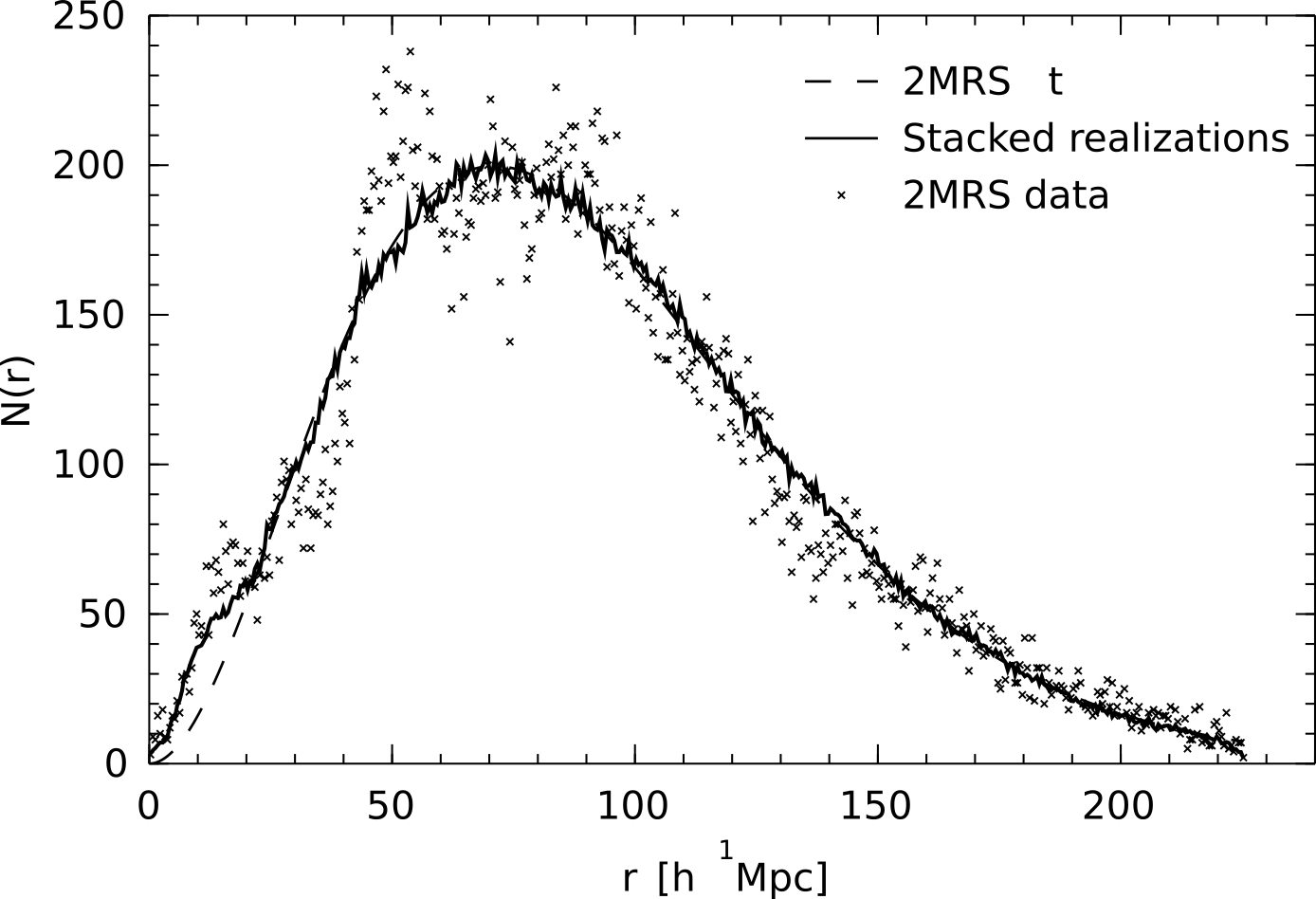} \caption{Fitting 2MRS $N(z)$ with the \textsc{sham} model. Dasehd curve corresponds to the absolute deviations fit of 2MRS data which is represented by the points in the figure. Solid curve corresponds to 30 stacked \textsc{sham} realizations.} \label{fig:FLUX_NZ} 
\end{figure}

\section{equations of motion} \label{sec:EQUATIONS_OF_MOTION}

We work within the standard paradigm in which gravity is the only long range force driving structure formation in the universe. Focusing on sufficiently large scales, we neglect gas physics and consider a single component of collision-less particles (mass tracers). The corresponding Euler equation is \citep{Peeb80} 
\begin{equation}
	\label{eq:NEWTONIAN_EQUATION_OF_MOTION} \dot{\vec{u}} + 2 H\vec{u} = - \vec{\nabla_x} \phi_g 
\end{equation}
where $\phi_g$ is the peculiar gravitational field given by the Poisson equation 
\begin{equation}
	\label{eq:PECULIAR_GRAVITATIONAL_POISSON_EQ} \nabla^2_x \phi_g = \frac{3}{2} \Omega H^2 a^3 \delta \; . 
\end{equation}
Further, the system obeys the continuity equation 
\begin{equation}
	\frac {d \ln (1+\delta)}{d t}=-\vec{\nabla_x} \cdot \vec {u}\; . 
\end{equation}
Initial particle positions and velocities can be advanced in time by solving these equations numerically using N-body simulations. These simulations have been instrumental in our understanding of gravitational dynamics and in the comparison between models and observations in terms of various statistical measures e.g. power spectra, correlation functions, topological properties etc. Augmented with realistic recipes for galaxy formation, the simulations are the standard platform for mimicking the data and providing robust error estimation. 

In addition to the full equations presented above, we will work with approximations thereof. These include linear theory \citep{Peeb80}, the Zel'dovich approximation and second order Lagrangian perturbation theory (2LPT), as described below. All results are benchmarked against linear theory. 

\subsection{Linear theory} This is strictly applicable in the limit of small density fluctuations, $|\delta| \ll 1$. To linear order, the equations above yield $\delta(\vec{x},t)=D^+(t) \delta^+(\vec{x})+D^-(t)\delta^-(\vec{x}) $ where $D^+$ and $D^-$ are the growing and decaying mode solutions to the equation 
\begin{equation}
	\label{eq:linearone} \ddot D +2H \dot D=\frac{3}{2} \Omega H^2 D\; . 
\end{equation}
By setting $\delta^-=0$, the past distribution of matter can be trivially derived from the density contrast $\delta_0(\vec{x})$ at the present time as $\delta_{\mathrm lin}=\delta_0(\vec{x})/D(t)$ where, hereafter, $D(t)$ denotes the growing mode normalized to unity at the present time. The omission of the decaying mode guarantees homogeneity at very early times $t\rightarrow 0$.

The density contrast can be written in terms of the divergence of a displacement field from initial to final positions as $\delta = \vec{\nabla} ( D\vec{\Psi} )$ where $\vec{\Psi}$ is independent of time and is a function of the particle initial (Lagrangian) coordinate only. Plugging this into the Poisson eq. \ref{eq:PECULIAR_GRAVITATIONAL_POISSON_EQ} gives 
\begin{equation}
	\label{eq:DISPLACEMENT_FIELD} \vec{\Psi} = -\frac{2 \vec{\nabla_x} \phi_g} {3 \Omega H^2 a^3 D} \; . 
\end{equation}
The linear comoving peculiar velocity is 
\begin{equation}
	\label{eq:LINEAR_COMOVING_VELOCITY} \dot\vec{x} = \dot{D} \vec{\Psi} = D H f \vec{\Psi} = -\frac{2 f \vec{\nabla_x} \phi_g} {3 \Omega H a^3} 
\end{equation}
where $f \equiv \rm{d}ln (D) / \rm{d} ln (a)$. Analytical approximations to $D$ and $f$ can be found in the literature, in this work we use the approximations by \cite{Bouchet95} $f=\Omega_M^{5/9}$ and \cite{Carroll92} 
\begin{equation}
	\label{eq:D_APPROX} D=\frac{2.5\Omega_m a}{\Omega^{4/7}-\Omega_{\Lambda}+\left( 1+\Omega_m /2 \right) \left( 1+\Omega_{\Lambda}/70 \right)} \; . 
\end{equation}

\subsection{2LPT and the Zel'dovich approximation} \label{sec:2LPT_ZELD_APPROX} The linear theory expression for the evolution of $\delta$ assumes that a patch of matter remains at its original position. In Lagrangian perturbation theory, the displacement is incorporated as an expansion in powers of $D$. In 2LPT the orbits are expressed as 
\begin{equation}
	\label{eq:2LPT_DYNAMICS} \vec{x(a)} = \vec{q} + D(a) \vec{\Psi} + D_2(a) \vec{\Psi^{(2)}}\; . 
\end{equation}
In this expression $\vec{q}$ is the initial (Lagrangian) comoving particle position, $\Psi$ is given in (\ref{eq:DISPLACEMENT_FIELD}) and 
\begin{equation}
	\label{eq:2ND_ORDER_DISPLACEMENT} \vec{\nabla} \vec{\Psi^{(2)}} = \frac{1}{2}\sum_{i \not= j}{\left( \vec{\Psi_{i,i}} \vec{\Psi_{j,j}} - \vec{\Psi_{i,j}} \vec{\Psi_{j,i}} \right)}\; , 
\end{equation}
where the indices $i,j$ represent Cartesian components. We use $f_2=2\Omega^{6/11}$ and $D=-3D_1^{2/7}$. The Zel'dovich approximation \citep{1970A&A.....5...84Z} is obtained by maintaining the term involving $\vec{\Psi}^{(2)}$, ie 
\begin{equation}
	\label{eq:ZELDOVICH_DYNAMICS} \vec{x(a)} = \vec{q}+ D(a) \vec{\Psi(\vec{q})}\; . 
\end{equation}

\section{ N{\scriptsize{o}}AM} \label{sec:THE_NOAM_METHOD}

Let $\vec{X}_{\mathrm obs}$ be the target positions, representing the observed distribution of galaxies (particles) at redshift $z=0$. Fix the initial redshift, $z_{\mathrm ini}$, corresponding to a sufficiently early time where linear theory is applicable. Let $\vec{Q}$ be particle positions at $z_{\mathrm ini}$ and $\vec{U}$ the corresponding linear peculiar velocities computed according to (\ref{eq:LINEAR_COMOVING_VELOCITY}). Starting from $\vec{Q}$ and $\vec{U}$ as initial conditions, NoAM runs the EoM (or any approximation thereof) to obtain the corresponding positions $\vec{X}$ at $z=0$. The algorithm searches for $\vec{Q}$ by minimizing 
\begin{equation}
	\label{eq:Cl} {\Delta}=L_p^{-1}\|\vec{X}_{\mathrm obs}-\vec{X} \| \; , 
\end{equation}
which is the RMS of the difference between the target and the (approximate) final positions in units of the mean particle separation, $L_p$. In practice, it suffices to reach $\vec{Q}$ which yields $\Delta<\rm Tol$ where $\rm Tol$ is a specified threshold. The minimization is achieved by means of an iterative procedure described below starting from $\tilde z_{\mathrm ini}=0$ and $\tilde \vec{Q}=\vec{X}_{\mathrm obs}$. 
\begin{enumerate}
	\item \label{item:RUN_SIM} advance the particles using the the EoM (or an approximation thereof) from $z=\tilde z_{\mathrm ini}$ to $z=0$ to obtain $\tilde \vec{X}$. 
	\item update $\tilde z_{\mathrm ini}\rightarrow \tilde z_{\mathrm ini} +\delta z$ where $\delta z$ is sufficiently small. 
	\item \label{item:UPDATEI} update $\tilde \vec{Q} \rightarrow \widetilde \vec{Q} + \lambda(\vec{X}_{\mathrm obs}-\tilde \vec{X}) $ and the corresponding $\vec{U}$ using relation \ref{eq:LINEAR_COMOVING_VELOCITY}. The numerical factor $\lambda\ll 1$ is introduced to stabilize the procedure. 
	\item goto \ref{item:RUN_SIM} until $\tilde z_{\mathrm ini}= z_{\mathrm ini}$. 
	\item end the iterations if $\Delta<\rm {Tol}$, otherwise, repeat steps \ref{item:RUN_SIM}--\ref{item:UPDATEI} until the criterion is satisfied. 
	\item identify $\vec{Q}=\tilde \vec{Q}$. 
\end{enumerate}
The algorithm can easily be implemented for any boundary conditions and approximation of the EoM. The method is efficient and trivial to adapt to \texttt{\bf s}-space data by replacing $\vec{T}$ and $\vec{F}$ with the corresponding redshift coordinates. In the applications to the mock catalogs below, NoAM is typicaly used with ${\rm Tol}=0.03$ and $z_{\mathrm ini}=50$.

\section{Mock Catalogs} \label{sec:MOCKS}

We resort to the Dark-Sky public Simulations \citep[][hereafter DSS]{2014arXiv1407.2600S} to generate mock catalogs. The DSS are 5 very large DM-only simulations of varying mass resolutions and (periodic) box sizes, among them the largest DM simulation to date, consisting of more than a trillion particles. The simulations are run for the $\Lambda$CDM cosmology with $\Omega_m=0.295, \Omega_b=0.0468, \Omega_{\Lambda}=0.705, h=0.688$, and $\sigma_8=0.835$. Halos in the simulations are provided by means of the \textsc{rockstar} algorithm \citep{2012ascl.soft10008B} which performs a friends-of-friends clustering on both configuration and momentum space, and gives a full tree of substructure. We generate two types of catalogs.

\subsection{Ideal catalogs \itype: } 

These are two catalogs extracted from the simulation $ds14\_g\_1600\_4096$ of $4096^3$ particles, each of mass $4.9\times10^9\smh$, in a periodic box of $1600\mpch$ on the side. The simulation was run from an initial redshift of $135$ until $z=0$. The catalogs include main (parent) halos with more than $20$ particles corresponding to a minimum halo mass of $9.8 \times 10^{10} \smh$. From the distribution of these halos, we employ the clouds-in-cells (CIC) algorithm with periodic boundary conditions to generate two density fields on a uniform $512^3$ grid embedded in the full simulation box, i.e. a $3.2\mpch$ grid cell size. The two catalogs correspond to assigning two choices of particle weights in the CIC procedure: 
\begin{description}
	\item [\textsc{mass\_weight}] where the density field is obtained by assigning halos weights proportional to their respective masses. 
	\item [\textsc{eq\_weight}] where uniform weights are given to to all halos. 
\end{description}
The density correlation function for the \textsc{eq\_weight} catalog is shown in Fig. \ref{fig:GRID_BAO}, note the strong \rm{BAO} signal at $100\mpch$. The other, \textsc{mass\_weight} catalog, yields a very similar correlation function. 

In the application of NoAM with full dynamics to these catalogs, the EoM are solved by means of a Particle-Mesh (PM) N-body code \citep{Eastwood74} with particles CIC-interpolated onto a periodic box of $512^3$ cells. The gravitational field is computed from the density field using Fast-Fourier Transform (FFT). A gaussian smoothing of radius $3\mpch$ is applied to the gravitational filed which is then CIC interpolated back to the particle positions. Each catalog contains $1.5\times 10^8$ particles (halos) and, for the N-body dynamics, their positions and velocities are integrated in time via a symplectic scheme \citep{1997astro.ph.10043Q}. We have compared the full N-body dynamics using the PM code to Gadget2 \citep{2005MNRAS.364.1105S}. The results are similar with small differences that could be explained by different smoothing and time integration schemes (e.g. Gadget uses spline smoothing and we use a Plummer potential \citep{1911MNRAS..71..460P}).

\subsection{ Realistic catalogs \rtype:} These are extracted from the simulation $ds14\_g\_800\_4096$ of $4096^3$ particles in a periodic box of $800\mpch$ on the side, corresponding to a particle mass of $6.1\times10^8\smh$. The simulation was run from redshift $z=183$ until today. We consider only halos containing no less than $50$ particles giving a minimal halo mass of $3.05 \times 10^{10} \smh$. For NoAM applications, the gravitational smoothing used with these catalogs depends on the particle mass, as described below.

We build 18 categories of realistic catalogs, with each category consisting of 30 random realizations. Each realization consists of particles within a spherical volume of radius $200\mpch$, the total mass of the particles is scaled to the mean mass of such sphere according to the DSS cosmological parameters. The gravitational smoothing, selection method, and specific particle masses all differ systematically between catalogs and is described below. All catalogs use the same set of locations as centers for mock realizations, these are selected as the center of a random halo (assuming periodic boundaries) with the following restrictions: 
\begin{enumerate*}
	\item the central halo is no less massive than $5\times10^{11}\smh$. 
	\item the velocity within shells at radii $3, 5,$ and $7\mpch$ doesn't differ by more than $50 \kms$. 
	\item no halo more massive than $10^{15}\smh$ is present within a distance of $20\mpch$. 
\end{enumerate*}
To minimize boundary effects only the inner $150\mpch$ of each realization are considered for velocity comparison. This sphere represents $\sim2.8\%$ of the simulation volume, and the cumulative volume of 30 realizations is similar to that of the simulation box. When comparing to 2MRS we use redshifted mocks (using the simulated ``real" particle velocity). The velocity used for comparison with reconstructions when using catalogs which include subhalos (except for the \textsc{subvel} catalog as discussed later), is the top halo velocity, i.e. the velocity of the top halo in the containing hierarchy (e.g. parent halo if the hierarchy is shallow) 
\begin{center}
	\begin{table}
		\centering 
		\begin{tabular}
			{ c S[table-format=-2.3, table-figures-uncertainty=1, table-number-alignment = center] S[table-format=-2.3, table-figures-uncertainty=1, table-number-alignment = center] } \hline Parameter & Halos & Subhalos \\
			\hline $M_1$ & 11.065 \pm 0.005 & 10.584 \pm 0.005\\
			$\epsilon$ & -4.623 \pm 0.002 & -4.874 \pm 0.002 \\
			$\alpha$ & -2.574 \pm 0.001 & -3.572 \pm 0.002 \\
			$\lambda$ & 6.590 \pm 0.003 & 2.531 \pm 0.002 \\
			$\gamma$ & 0.166 \pm 0.001 & 0.084 \pm 0.001 \\
			$F_{th}$ & 18894 \pm 15 & 18894 \pm 15 \\
			\hline 
		\end{tabular}
		\caption{SHAM model parameters.} \label{table:sham_parameters} 
	\end{table}
\end{center}

\begin{figure*}
	\includegraphics[width=462pt]{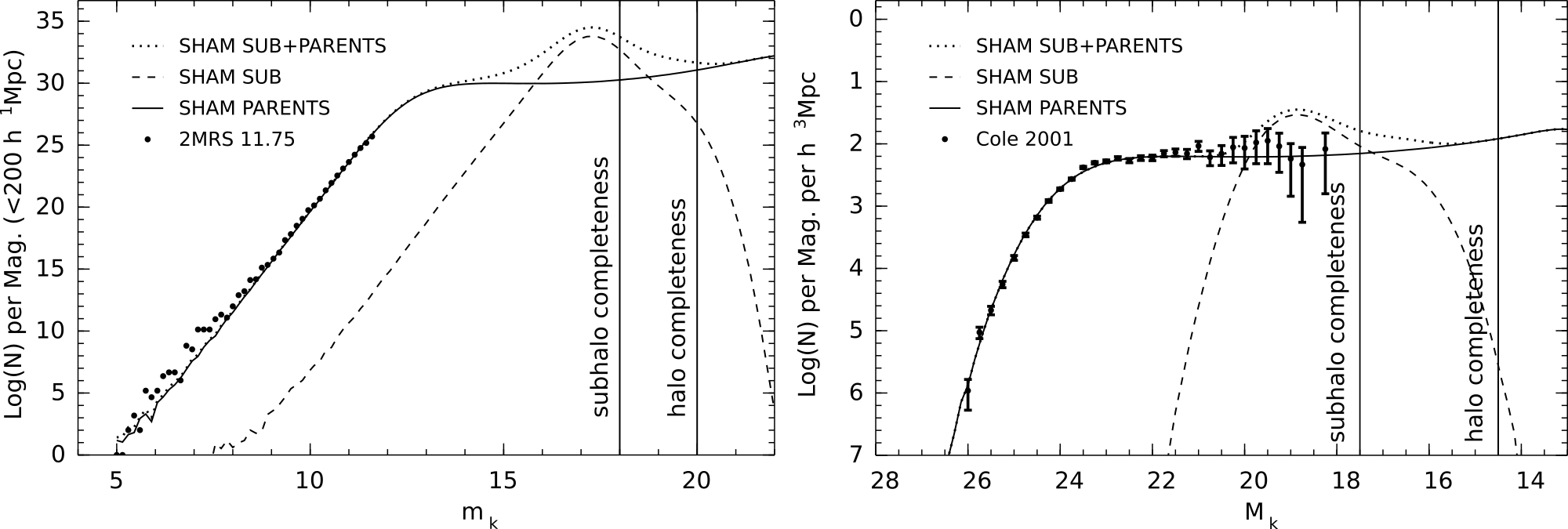} \caption{Comparison of \textsc{sham} catalog luminosity functions to observations. {\it Left:} compared to 2MRS apparent isophotal $k$-magnitudes {\it Right:} compared to absolute $k$-magnitudes (${\rm K_{ron}}$). Indicated also are $95\%$ completeness limits as estimated by a simple fit to the mass histogram of halos contributing to each magnitude bin. } \label{fig:lum_fit} 
\end{figure*}

All realistic mock catalogs (except the volume limited ones) require as input a smooth, physically driven fit to the 2RMS number function $N(z)$. For this purpose we parametrize the number of galaxies in $0.5\mpch$ thick shells as 
\begin{equation}
	\label{eq:2MRS} N_{fit}(r) = \alpha r^\beta e^{-(r/\gamma)^\delta} 
\end{equation}
where $r$ is the shell radius. We fit between $20$ to $225\mpch$ by minimizing the sum of absolute deviations 
\begin{equation}
	\label{eq:2MRS_POISSONIC_DISTANCE} \sum_{i}{\frac{\left| N_{real}(r_i)-N_{fit}(r_i) \right|}{N_{fit}(r_i)}} 
\end{equation}
over all shells $i$. The minimization gives $\alpha=0.193, \beta=1.961, \gamma=55.062$, and $\delta=1.38$. This is our baseline fit; it gives a mean number of particles only $0.75\%$ higher than what 2MRS contains in the fitted region. In contrast using least squared deviations leads to a mean particle number $4.5\%$ lower. We test the sensitivity of our results to systematics of $N(z)$ (e.g. the effects of different fitting methods) by scaling $N(z)$ by $\pm10\%$; see Fig. \ref{fig:2mrs_nz} where the solid curve represents a fit by absolute deviations, the dashed curve by squared deviations, and dotted curve is the 2MRS data. Grey areas represent the scaled $N(z)$ (the fit by squared deviations is included in this region). The catalog names are listed in Table \ref{table:main_results}, according to the following key: 
\begin{description}
	\item [\textsc{vol}] Volume limited. These realizations consist of halos with a minimum mass of $1.5\times10^{11}\smh$. The mean number of particles (halos) inside a $150\mpch$ sphere radius of $~45000$ halos, roughly similar to that contained in the 2MRS. The particles are given equal weight independent of their mass and the gravitational smoothing is set to $3\mpch$.
	
	\item [\textsc{particles}] Volume limited particle realizations. Number of particles, mass and gravitational smoothing are the same as above.
	
	\item [\textsc{flux}] Flux limited. The number of halos at each $0.5\mpch$ thick shell is chosen as a Poisson random variable with frequency 
	\begin{equation}
		\label{eq:FLUX_N_IN_SHELL} \lambda(i)=N_{fit}(i) \frac{N_{sim}(i)}{V_{sim}(i)} \left( \frac{N_{sim}}{V_{sim}} \right) ^{-1} 
	\end{equation}
	where $N_{sim}(i)$ is the number of halos in the simulation in shell $i$, $V_{sim}(i)$ is the volume of the shell, $N_{sim}$ is the total number of halos in the simulations and $V_{sim}$ is the simulation box volume. Each shell is given a total mass corresponding to its volume times the mean universe density, and this mass is equally divided by all the particles in the shell. In the simulation, the shell contains many more halos than what is required by the above Poisson process. We employ two strategies to choose which halos to include out of the available ones, these strategies are described in the \textsc{mm} and \textsc{rm} tags below. In this realizations the gravitational smoothing length of particles depends on their mass as follows: 
	\begin{equation}
		\label{eq:FLUX_SMOOTHING} \epsilon= 58.87 \left(\frac{m_p}{10^{10}\smh} \right)^{2/3} \mpch\;. 
	\end{equation}
	
	\item [\textsc{mm}] This is a strategy for choosing halos in shells. When using this strategy shells are populated with most massive halos. This strategy has the strongest possible dependence on mass.
	
	\item [\textsc{rm}] This is another strategy for choosing halos in shells. When using this strategy shells are populated with random halos. This strategy has the minimal possible dependence on mass.
	
	\item [\textsc{s}] Subhalos are considered for populating the shells.
	
	\item [\textsc{ns}] Subhalos are not considered for populating the shells.
	
	\item [\textsc{ten}] Velocity field smoothed with a \texttt{\bf r}-space top-hat of radius of $10\mpch$ instead of $5\mpch$ (as is for the other catalogs).
	
	\item [\textsc{subvel}] The catalog uses subhalo velocities for comparison with reconstruction. Other catalogs use the top containing halo velocity, i.e. the top halo in the containing hierarchy.
	
	\item [\textsc{sham}] Flux limited realizations. Shells are populated by choosing all simulated galaxies with a flux above a threshold $F_{th}$, corresponding to the 2MRS 11.75 $k$-magnitude limit. In the SHAM process each halo is assumed to harbor a galaxy, the galaxy luminosity $L$ follows a Mean Luminosity Function (MLF, denoted as $L(M_h)$) which describes the mean galactic luminosity as function of the host halo mass. Variance in this relation is modeled with a Poisson noise of $0.2$ dex. We use separate MLFs for halos and subhalos \citep[see][]{2013ApJ...767...92R}, each parametrized with $5$ parameters as done by \cite{2010ApJ...717..379B}: 
	\begin{equation}
		\label{eq:FLUX_FIT} \log L(M_h) = \log(\epsilon M_1)+f(\log(M_h/M_1))-f(0) 
	\end{equation}
	where 
	\begin{equation}
		\label{eq:FLUX_FIT_F} f(x) = \delta \frac{\left( \log(1+e^x) \right)^\gamma}{1+e^{10^{-x}}}-\log(10^{ax}+1) \; . 
	\end{equation}
	
	The model has in total 11 parameters, we tune them by matching the 2MRS $N(z)$ with 30 stacked SHAM realizations. Parameter values are presented in Table \ref{table:sham_parameters} and the fitted $N(z)$ in Fig. \ref{fig:FLUX_NZ}. We use natural units for all model parameters. As a test for this SHAM model, we compare the resulting magnitudes to those observed in the 2MRS: The apparent $k$-magnitude is defined as ${\rm m}_k=-2.5 \log(F/F_0)=-2.5 \log(F)+2.5\log(F_0)$ where $F=L/(4\pi \/r^2)$ is the galaxy flux, $L$ the luminosity, $r$ the distance to the galaxy and $F_0$ the reference flux of the 2MRS photometric filter. Here we compare to the isophotal magnitudes (these are the primary 2MRS magnitudes). We interpret $F_{th}$ as the 2MRS threshold of $11.75$, i.e. ${\rm m}_{th}=11.75=-2.5 \log(F_{th})+2.5 \log(F_0)$. This gives $2.5 \log(F_0)=22.43$ hence ${\rm m}_k = 22.43 - 2.5 \log(F)$ for any SHAM flux $F$. The resulting fit is presented in the left panel of Fig. \ref{fig:lum_fit}. For the absolute $k$-magnitude we compare with the results of \cite{2001MNRAS.326..255C} (see their table 4) which used the 2MASS Kron magnitudes. We repeat the above exercise this time we tune $F_0$ so our results fit the observations and get ${\rm M}_k = -2.61 - 2.5 \log(L)$. The fit is presented in the right panel of Fig. \ref{fig:lum_fit}. Galaxy masses are given in the same way as for the flux limited catalogs, i.e. shell volume times mean universe density divided by number of particles in the shell. 
\end{description}

Fig. \ref{fig:REALIZATIONS_SCATTER} compares the scatter of galaxies in the 2MRS to several realizations at noted in the panels. While all catalogs agree on the 2MRS $N(z)$ (see Fig. {\ref{fig:nz_realizations}), and the SHAM ones agree on the luminosity function (see Fig. \ref{fig:FLUX_NZ}), it is evident that no method used in this study reproduces a catalog with 2MRS-like Fingers Of God (FOG). The \texttt{S\char`_MM} catalogs (with substructure and selection according to maximal posible mass) show too strong FOG, and oter catalogs with substructure show too little. Matching power spectrum at scale of FOG with SHAM is being actively researched, the most promising methods use halo properties at the time of accretion \citep[e.g.][]{2015arXiv150303973Y}. This methods cannot be used with the Dark Sky Simulations since the only available snapshots are at redshift zero. We do acknowledge that for such large simulations, where each snapshot is tens of terabytes, it is a challenge to make such data accessible to the community.

The gravitational field for the realistic catalogs is computed using a tree method \citep{1986Natur.324..446B} with mean density boundary conditions. The method approximates the following direct sum for each particle (particles enumerated by $i$) 
\begin{equation}
	\label{eq:DIRECT_GRAV_FIELD} \left(\vec{\nabla_x}\phi_g \right)_i = \frac{1}{a^3} \left[ G \sum_j\frac{m_j\vec{x_{ij}}}{p_{ij}^{3}} +\frac{1}{2}\Omega_0 H_0^2 \vec{x_i} \right] 
\end{equation}
where $p_{ij}=(x_{ij}^2+\epsilon^2)^{1/2}$ and $\epsilon$ is the standard Plummer sogtening. The cell opening criterion used is $0.5$.

Time integration for the full N-body dynamics uses the same symplectic scheme as the \itype catalogs. We tested the full N-body dynamics by comparing a simulation to Gadget2 \citep{2005MNRAS.364.1105S} and the results were similar with only small difference. Further, comparing linear velocity (as calculated by our tree code) to Dark-Sky velocities gives a good agreement, see Fig. \ref{fig:V_VS_V}. 
\begin{center}
	\begin{table*}
		\begin{tabular}
			{lccccccccc} \hline catalog & particles & lin & lin, S & Zel & Zel, S & 2LPT & 2LPT, S & dyn & dyn, S \\
			name & $( \times 1000)$ & $(\kms)$ & $(\kms)$ & $(\kms)$ & $(\kms)$ & $(\kms)$ & $(\kms)$ & $(\kms)$ & $(\kms)$ \\
			\hline \textsc{particles} & $44.3 \pm 2.1$ & $137 \pm 11$ & $201 \pm 12$ & $129 \pm 9$ & $200 \pm 11$ & $128 \pm 9$ & $201 \pm 11$ & $128 \pm 9$ & $196 \pm 11$\\
			\textsc{particles\_10} & $44.3 \pm 2.1$ & $101 \pm 14$ & $128 \pm 13$ & $94 \pm 12$ & $127 \pm 11$ & $94 \pm 12$ & $128 \pm 12$ & - & - \\
			\textsc{particles\_xl} & $442.6 \pm 14.0$ & $86 \pm 14$ & $161 \pm 13$ & $75 \pm 12$ & $161 \pm 12$ & $74 \pm 12$ & $162 \pm 12$ & $74 \pm 10$ & - \\
			\textsc{particles\_xxl} & $1842.1 \pm 58.2$ & $81 \pm 17$ & - & - & - & - & - & - & - \\
			\textsc{s\_lum\_mm} & $37.8 \pm 2.8$ & $134 \pm 15$ & $152 \pm 13$ & $133 \pm 16$ & $152 \pm 14$ & $133 \pm 16$ & $152 \pm 14$ & $134 \pm 15$ & $154 \pm 14$\\
			\textsc{s\_lum\_mm\_10} & $37.8 \pm 2.8$ & $112 \pm 17$ & $116 \pm 14$ & $112 \pm 17$ & $115 \pm 15$ & $111 \pm 17$ & $115 \pm 15$ & - & - \\
			\textsc{s\_lum\_mm\_xl} & $41.7 \pm 3.1$ & - & $150 \pm 13$ & - & - & - & - & - & - \\
			\textsc{s\_lum\_mm\_small} & $34.1 \pm 2.5$ & - & $154 \pm 13$ & - & - & - & - & - & - \\
			\textsc{s\_lum\_mm\_subvel} & $37.7 \pm 1.8$ & $240 \pm 20$ & $279 \pm 26$ & - & - & - & - & - & - \\
			\textsc{s\_lum\_rm} & $37.8 \pm 2.8$ & $126 \pm 12$ & $143 \pm 17$ & - & - & - & - & - & $143 \pm 16$\\
			\textsc{s\_lum\_rm\_xl} & $41.6 \pm 3.1$ & - & $141 \pm 16$ & - & - & - & - & - & - \\
			\textsc{s\_lum\_rm\_small} & $34.1 \pm 2.5$ & - & $144 \pm 17$ & - & - & - & - & - & - \\
			\textsc{ns\_lum\_mm} & $37.9 \pm 1.4$ & $125 \pm 14$ & $126 \pm 11$ & $130 \pm 14$ & $127 \pm 10$ & $128 \pm 14$ & $126 \pm 10$ & $128 \pm 14$ & $130 \pm 11$\\
			\textsc{ns\_lum\_rm} & $37.8 \pm 1.4$ & - & $145 \pm 10$ & - & - & - & - & - & - \\
			\textsc{ns\_lum\_rm\_xl} & $41.6 \pm 1.7$ & - & $143 \pm 14$ & - & - & - & - & - & - \\
			\textsc{ns\_lum\_rm\_small} & $34.0 \pm 1.3$ & - & $148 \pm 12$ & - & - & - & - & - & - \\
			\textsc{ns\_vol} & $43.3 \pm 2.0$ & $120 \pm 9$ & $129 \pm 10$ & $123 \pm 9$ & $133 \pm 9$ & $122 \pm 9$ & $133 \pm 9$ & $121 \pm 9$ & $129 \pm 9$ \\
			\textsc{sham} & $38.8 \pm 2.8$ & - & $138 \pm 16$ & - & - & - & - & - & - \\
			\hline 
		\end{tabular}
		\caption{\rtype catalogs reconstruction performance (for the cartesian X component). Catalog names, number of particles inside $150\mpch$, and the reconstruction performance of each method. Mean $\sigma$ from the different realizations for each catalog are presented, both for \texttt{\bf r} and \texttt{\bf s}-space. A scatter plot of the reconstructed versus true velocities is given in Fig. \ref{fig:V_VS_V}} \label{table:main_results} 
	\end{table*}
\end{center}
\begin{center}
	\begin{table*}
		\begin{tabular}
			{lccccccccc} \hline catalog name & lin & lin, S & Zel & Zel, S & 2LPT & 2LPT, S & dyn & dyn, S \\
			\hline \textsc{particles} & $0.92 \pm 0.04$& $0.78 \pm 0.03$ & $0.76 \pm 0.34$ & $0.75 \pm 0.13$ & $1.77 \pm 0.48$ & $1.47 \pm 0.38$ & $0.83 \pm 0.24$ & $0.75 \pm 0.23$\\
			\textsc{particles\_10} & $0.95 \pm 0.05$& $0.87 \pm 0.06$ & $0.77 \pm 0.34$ & $0.8 \pm 0.14$ & $1.83 \pm 0.51$ & $1.43 \pm 0.37$ & - & - \\
			\textsc{particles\_xl} & $0.97 \pm 0.05$& $0.84 \pm 0.05$ & $0.61 \pm 0.45$ & $0.71 \pm 0.25$ & $1.96 \pm 0.53$ & $1.39 \pm 0.37$ & $0.92 \pm 0.27$ & - \\
			\textsc{particles\_xxl} & $0.96 \pm 0.05$& - & - & - & - & - & - & - \\
			\textsc{s\_lum\_mm} & $0.67 \pm 0.06$& $0.56 \pm 0.05$ & $0.51 \pm 0.14$ & $0.42 \pm 0.11$ & $1.17 \pm 0.29$ & $0.96 \pm 0.24$ & $0.6 \pm 0.13$ & $0.5 \pm 0.12$\\
			\textsc{s\_lum\_mm\_10} & $0.65 \pm 0.07$& $0.57 \pm 0.06$ & $0.49 \pm 0.14$ & $0.42 \pm 0.11$ & $1.14 \pm 0.29$ & $0.96 \pm 0.25$ & - & - \\
			\textsc{s\_lum\_mm\_xl} & - & $0.57 \pm 0.054$ & - & - & - & - & - & - \\
			\textsc{s\_lum\_mm\_small} & - & $0.55 \pm 0.05$ & - & - & - & - & - & - \\
			\textsc{s\_lum\_mm\_subvel} & $0.74 \pm 0.08$& $0.43 \pm 0.06$ & - & - & - & - & - & - \\
			\textsc{s\_lum\_rm} & $1.06 \pm 0.07$& $0.89 \pm 0.1$ & - & - & - & - & - & $0.79 \pm 0.22$\\
			\textsc{s\_lum\_rm\_xl} & - & $0.89 \pm 0.1$ & - & - & - & - & - & - \\
			\textsc{s\_lum\_rm\_small} & - & $0.89 \pm 0.086$ & - & - & - & - & - & - \\
			\textsc{ns\_lum\_mm} & $0.96 \pm 0.08$& $0.88 \pm 0.08$ & $0.69 \pm 0.16$ & $0.59 \pm 0.15$ & $1.63 \pm 0.42$ & $1.41 \pm 0.37$ & $0.83 \pm 0.2$ & $0.74 \pm 0.19$\\
			\textsc{ns\_lum\_rm} & - & $1.15 \pm 0.12$ & - & - & - & - & - & - \\
			\textsc{ns\_lum\_rm\_xl} & - & $1.13 \pm 0.12$ & - & - & - & - & - & - \\
			\textsc{ns\_lum\_rm\_small} & - & $1.13 \pm 0.12$ & - & - & - & - & - & - \\
			\textsc{ns\_vol} & $0.98 \pm 0.07$& $0.91 \pm 0.07$ & $0.7 \pm 0.3$ & $0.77 \pm 0.14$ & $1.77 \pm 0.47$ & $1.47 \pm 0.37$ & $0.87 \pm 0.24$ & $0.82 \pm 0.23$ \\
			\textsc{sham} & - & $0.86 \pm 0.11$ & - & - & - & - & - & - \\
			\hline 
		\end{tabular}
		\caption{\rtype catalogs reconstruction performance (for the cartesian X component). Mean slope from the different realizations for each catalog are presented, both for \texttt{\bf r} and \texttt{\bf s}-space.} \label{table:main_results_slope} 
	\end{table*}
\end{center}
\begin{center}
	\begin{table*}
		\begin{tabular}
			{lccccccccc} \hline catalog name & lin $\sigma$ & lin slope & Zel $\sigma$ & Zel slope & 2LPT & 2LPT slope & dyn $\sigma$ & dyn slope \\
			& $(\kms)$ & & $(\kms)$ & & $(\kms)$ & & $(\kms)$ & \\
			\hline \textsc{particles} & $138 \pm 11$ & $0.92 \pm 0.04$ & $129 \pm 10$ & $0.77 \pm 0.3$ & $129 \pm 10$ & $1.78 \pm 0.45$ & $129 \pm 10$ & $0.84 \pm 0.24$ \\
			\textsc{s\_lum\_mm} & $135 \pm 16$ & $0.68 \pm 0.07$ & $133 \pm 16$ & $0.51 \pm 0.15$ & $134 \pm 16$ & $1.15 \pm 0.29$ & $134 \pm 16$ & $0.59 \pm 0.13$ \\
			\hline 
		\end{tabular}
		\caption{\rtype catalogs radial reconstruction performance} \label{table:main_results_radial} 
	\end{table*}
\end{center}
\begin{figure*}
	\includegraphics[width=462pt]{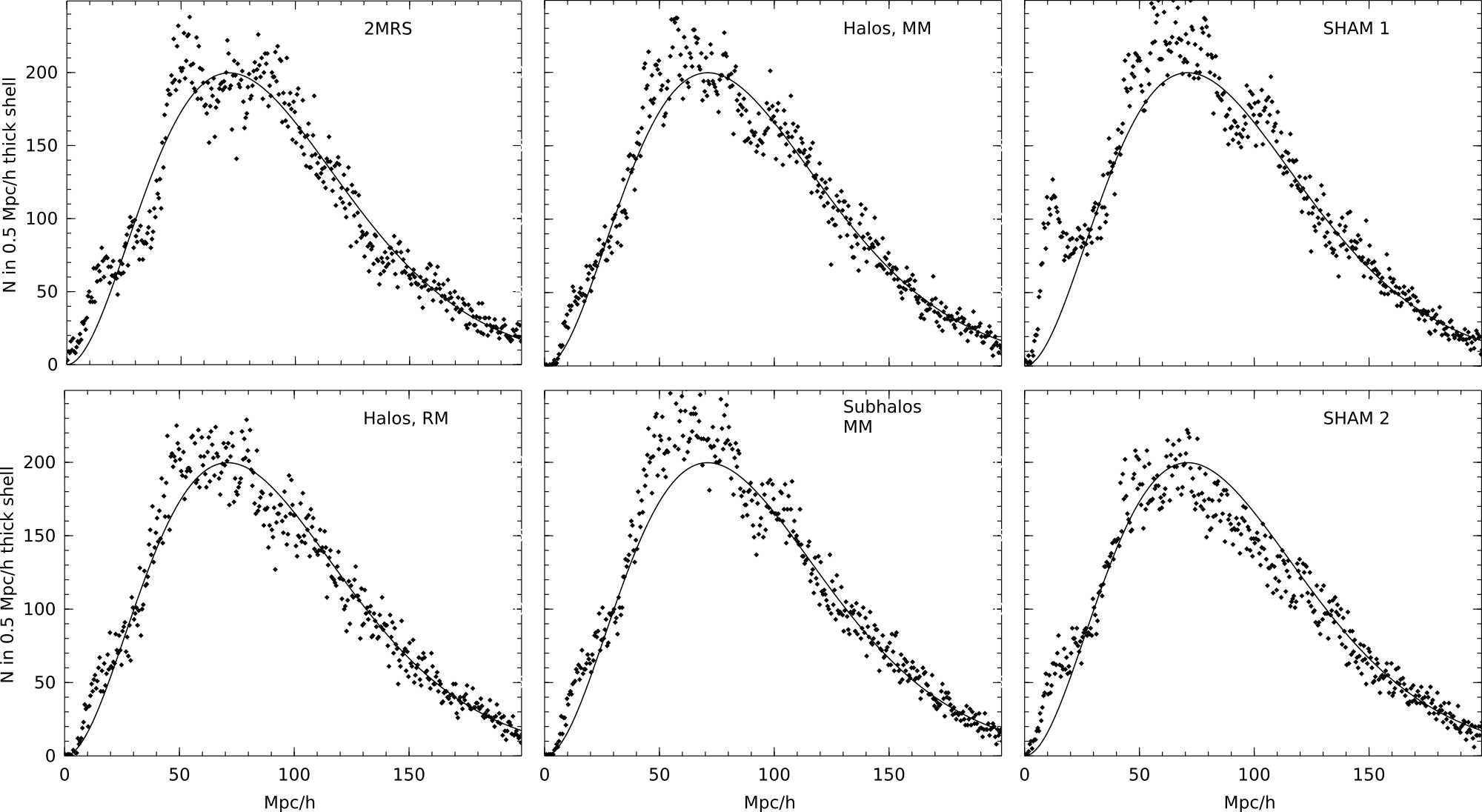} \caption{$N(z)$ in bins of $0.5\mpch$ thick spherical shells, as function of shell radius, for specific realizations of catalogs as denoted in the panels. Top left panel corresponds to measured 2MRS catalog.} \label{fig:nz_realizations} 
\end{figure*}
\begin{figure*}
	\includegraphics[width=14cm]{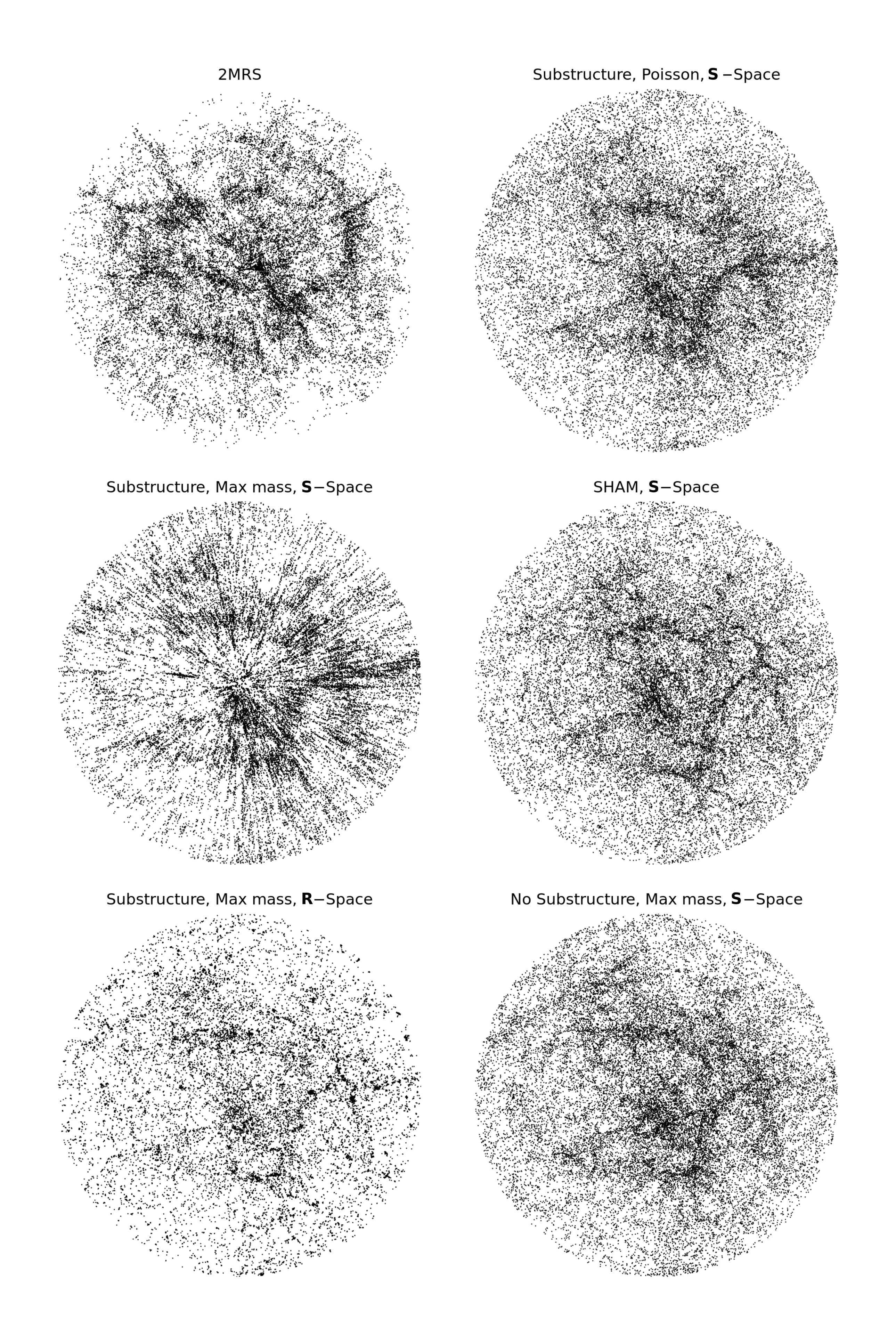} \caption{Scatter plot of realizations of several catalogs as indicated in the panels. Top left image corresponds to the measured 2MRS.} \label{fig:REALIZATIONS_SCATTER} 
\end{figure*}
\begin{figure*}
	\includegraphics[width=462px]{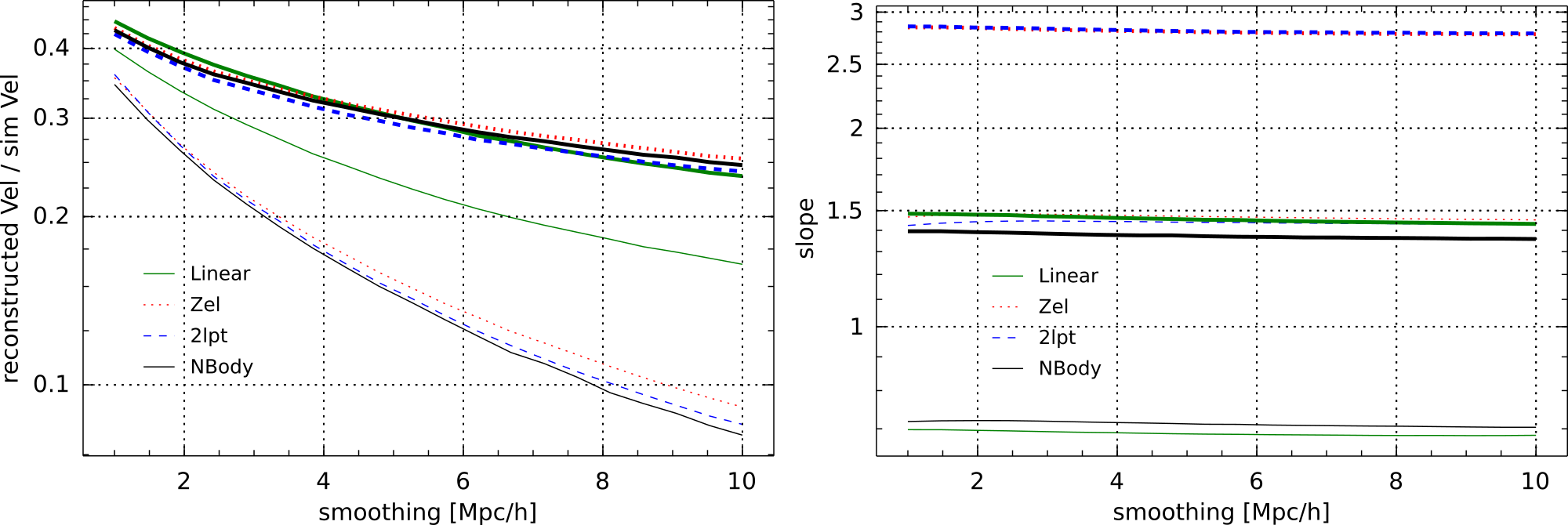} \caption{{\it left: } Ratio of reconstructed to true $\sigma_v$ as function of smoothing (tophat in \texttt{\bf r}-space), calculated on a grid for the \itype catalogs. Thick and thin curves correspond to \textsc{eq\_weight} and \textsc{mass\_weight} catalogs. Reconstruction type is indicated in the legend. {\it right: }slope of velocity reconstruction as function of smoothing for the same curves as in the left panel} \label{fig:gird_vel_ratio} 
\end{figure*}
\begin{figure*}
	\includegraphics[width=16.7cm]{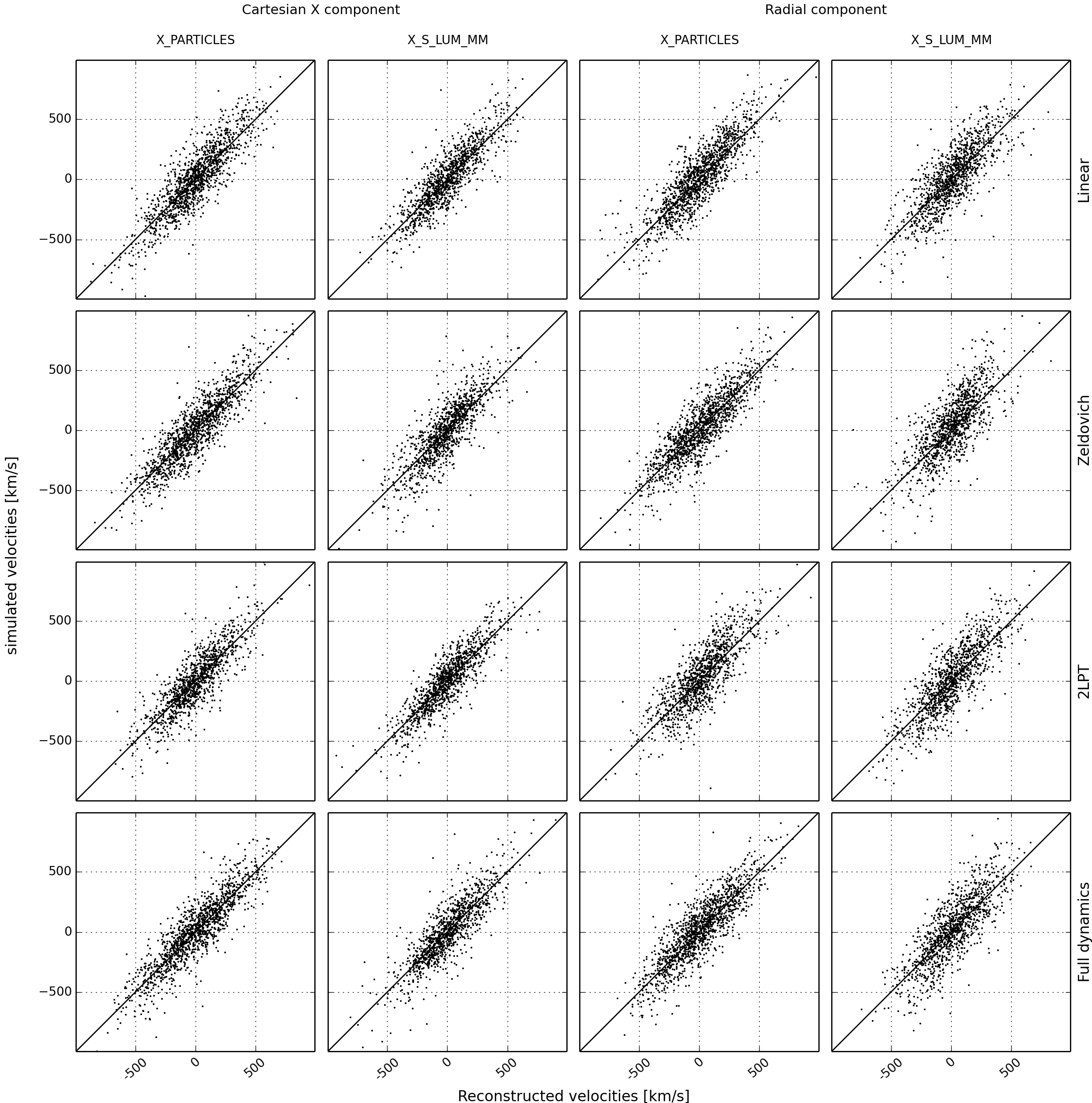} \caption{X and radial components of $\vrec$ vs $\vtru$ for two \rtype catalogs (labeled at the top). All reconstructions (indicated to the right) are in \texttt{\bf r}-space. In each panel a random subset of $0.1\%$ of all particles in all 30 realizations are plotted. Reconstructed velocities are top-hat filtered over $5\mpch$ and scaled so the the slope of the regression of $\vtru$ on $\vrec$ in the figure is unity. The un-scaled slopes are given in Table \ref{table:main_results_slope} and the dispersion in Table \ref{table:main_results}} \label{fig:V_VS_V} 
\end{figure*}

\section{Results} \label{sec:RESULTS}

We have applied NoAM to the catalogs using the full dynamics, the Zel'dovich and 2LPT approximations, with the parameters given in \S \ref{sec:THE_NOAM_METHOD}. For the \rtype catalogs the reconstruction is performed in both \texttt{\bf r} and \texttt{\bf s}-space, while only \texttt{\bf r}-space is considered for the \itype catalogs. All results are compared to the linear theory velocity given in terms of the gravity field of the observed distribution according to relation \ref{eq:LINEAR_COMOVING_VELOCITY}. For linear reconstruction in \texttt{\bf s}-space we use an iterative method \citep{2002MNRAS.335...53B} to achieve self-consistent density and velocity fields. We denote by $\vrec$ and $\vtru$ the reconstructed and true velocities (as simulated in the DSS) for all tracers (halos). Further, $\vrrec$, $\vrtru$ and $\vxrec$, $\vxtru$ are the radial and cartesian X components of the velocity, respectively. The comparison between $\vrec$ and $\vtru$ is performed after smoothing of these velocities with the same top-hat filter. The smoothing is essential in order for the comparison to assess the same physical scales and also to eliminate small scale nonlinearities (e.g. incoherence velocities) which are not reproduced in the reconstruction. To quantify the quality of the reconstructed velocities in each case we compute $\sigma_{{\delta v}x}=\langle(b\vxrec-\vxtru)^2\rangle^{1/2}$, $\sigma_{{\delta v}r}=\langle(b\vrrec-\vrtru)^2\rangle^{1/2}$ and compare to $\sigma_x=\langle(\vxtru)^2\rangle^{1/2}$ and $\sigma_r=\langle(\vrtru)^2\rangle^{1/2}$ respectively. Here $b$ is a linear proportionality factor (slope) which minimizes $\sigma_{\delta v}$. For the \itype catalogs, only the cartesian X component is considered because of the lack of an ``observer" defining a center. Results for velocity reconstructions of these catalogs are summarized in Fig. \ref{fig:gird_vel_ratio}. The left panel shows the ratio $\sigma_{\delta v}/\sigma$ versus the smoothing length (mean particle separation is $3\mpch$). Mass weighting (thin curve) yields a significantly better agreement between $\vrec$ and $\vtru$: with smoothing of $8\mpch$, the \rm{2LPT} reconstruction gives $\sigma_{\delta v}/\sigma \sim 10\%$ and $\sim 25\%$, respectively, for \textsc{mass\_weight} and \textsc{eq\_weight} catalogs. Linear theory with mass weighting performs poorly compared to the non-linear reconstructions where the latter all yield similar results. For uniform weighting, the performance of all reconstructions is comparable. The panel to the right of Fig. \ref{fig:gird_vel_ratio} shows the slope as a function of the smoothing length for the \itype catalogs. This slope depends on the reconstruction method used, and the less sensitive it is to the catalog type (or details) the more confident we can be regarding systematic errors in the application to real observed catalogs. We emphasize that all reconstructions are performed assuming that particles in the mocks are unbiased relative to the dark matter. Therefore, the slopes are expected to deviate from unity. For reference, the linear biasing factor appropriate for halos of mass $M\sim 5\times 10^{10} h^{-1} M_\odot$ is $b_h= 0.8$ \citep{2001MNRAS.323....1S}

We focus now on \rtype catalogs. Fig. \ref{fig:V_VS_V} shows a scatter plot comparison of $\vtru$ vs $\vrec$ velocities in \texttt{\bf r}-space, for 2 \rtype catalogs, as indicated in the figure. The performance of the reconstruction for the \rtype catalogs is quantified in tables \ref{table:main_results} and \ref{table:main_results_slope} for the X-cartesian component of velocities. Results for the radial velocity components are listed in Table \ref{table:main_results_radial}. The quality of the reconstructed radial and X components does not differ significantly and we continue our analysis using the cartesian component only. In agreement with the results of the \itype catalogs, linear reconstruction shows comparatively low performance only for the ``high information" particle catalogs. While linear reconstruction gives $\sigma=137\kms$ (in \texttt{\bf r}-space), NoAM with N-body dynamics gives $\sigma=128\kms$. This difference is even more pronounced for the $\rm{XL}$ catalog of particles where linear theory gives $\sigma=86\kms$ while the N-body dynamics yields $\sigma=74\kms$. For other \texttt{\bf r}-space catalogs and for all catalogs in \texttt{\bf s}-space, all methods show similar performance. Cosmic variance is responsible for $\pm \sim 10\kms$, again similar for all methods. In contrast to the results of the \itype catalogs, the linear reconstruction slope is closer to unity than non-linear methods. Further, it is more stable across the various catalogs. 

Enhancement of the BAO signal for the \itype \textsc{mass\_weight} catalog is demonstrated in Fig. \ref{fig:GRID_BAO}. The solid black curve is the density correlation function computed from the $z=0$ simulation output, other curves correspond to back-in-time configurations obtained from different dynamical reconstructions, as indicated in the figure. In the ``backwards Zel'dovich", the particles are moved in the opposite direction to their linear velocity at redshift $z=0$. For the Zel'dovich dynamics we move the particles backward according to their Zel'dovich reconstruction velocity. The mean $\rm{1D}$ distance maximizing the BAO signal for both methods is $1.3\mpch$. For the N-body dynamics we run the simulation from the initial redshift of $50$ to a final redshift $0.7$ (this maximizes the BAO signal). All methods enhance the BAO signal, and we find no significant difference between the different methods. The signal enhancement is in line with previous results \citep[e.g.][]{Kazin2014}. Interestingly, \cite{2015PhRvD..92h3523A} find a slight 2LPT advantage over Zel'dovich for BAO signal reconstruction, but they use a different smoothing scheme for each method ($10\mpch$ for Zel'dovich and $5 - 7.5\mpch$ for 2LPT, c.f. their Fig. 14). In both works the basic smoothing is similar to the mean particle separation ($12\mpch$ in \cite{2015PhRvD..92h3523A} and $3\mpch$ for the \itype catalogs). It is not clear in their paper how Zel'dovich compares with a higher order scheme when using smaller smoothing lengths. 
\begin{figure}
	\includegraphics[width=8.1cm]{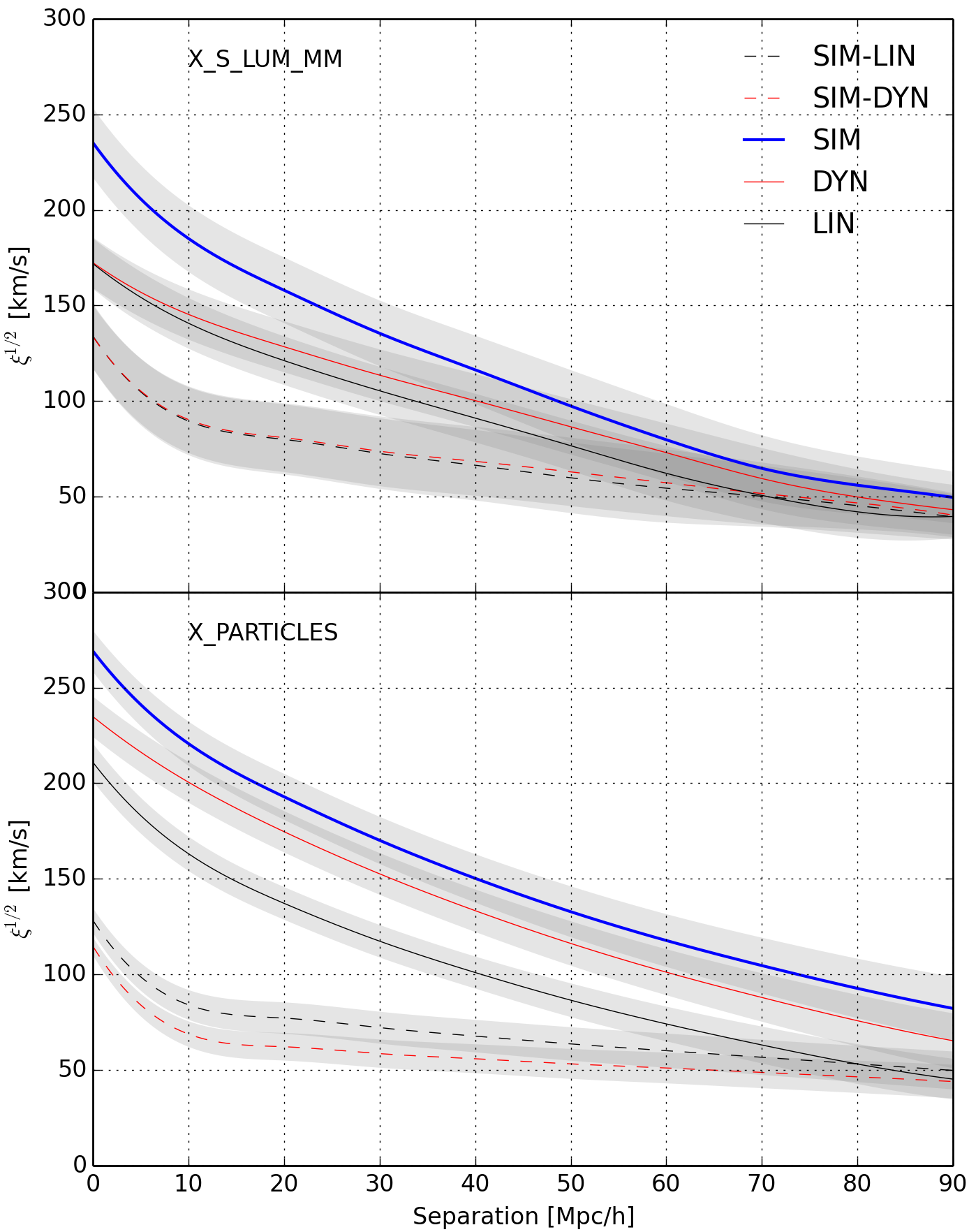} \caption{Radial velocity correlations for two \rtype catalogs, as indicated in the figure. Solid curves correspond to $\vtru$ in the simulation and $\vrec$ from two reconstructions, dynamical and linear. Dashed curved are computed from the residuals $\vtru-\vrec$. Grey areas represent half the standard deviation among all 30 realizations per catalog. } \label{fig:REALISTIC_CORR} 
\end{figure}

Figure \ref{fig:REALISTIC_CORR} shows the velocity correlation function for two \rtype catalogs (volume limited particles and a flux Limited galaxies, as indicated in the figure). This correlation is defined as \citep{Gorski89} 
\begin{equation}
	\label{eq:VEL_CORR} \Psi = \frac{ \sum_{i\neq j} {v}_{i r} {v}_{j r} \cos{\theta_{ij}}^2} {\sum_{i\neq j} \cos{\theta_{ij}}^2} 
\end{equation}
where $v_{r}$ is the radial peculiar velocity of a particle and and $\theta_{ij}$ is the angle between the lines of sight to two particles. Solid blue heavy curves represent correlation of true velocities, while solid red and black curves represent reconstructed velocities using full dynamics and Zel'dovich respectively. Dashed curves represent the residual velocity $v^{res}_r \equiv v^{tru}_r-v^{rec}_r$. For the catalog \textsc{particles} the full dynamics reconstruction achieves a lower residual, albeit in agreement with other results in this paper this advantage disappears in the flux limited catalog, \textsc{s\_mm}.

Linear reconstruction performs surprisingly well. Previous studies comparing velocity or initial conditions reconstructions using ideal catalogs find, in agreement with this study, either very good linear performance (e.g when dealing with expansion of cosmic voids, see \cite{2014PhRvL.112y1302H}) or poor linear performance \citep[e.g.][]{Nusser2000, 1997MNRAS.285..793C}, and better systematic behaviour for Zel'dovich (or 2LPT) dynamics. However, regarding the latter case, we find that other methods fall short of expectations when applied to more realistic \rtype catalogs, and in practice perform worse than linear for both systematic and inherent errors. This is especially important for catalogs with volumes not large enough to cancel cosmic variability as the linear reconstruction will have significant smaller systematic errors.

For the reasons stated above, we reconstruct the 2MRS velocity field using a linear method. We choose the slope of the linear reconstruction which minimizes the total random error in the reconstructed velocity, considering all of the \rtype catalogs in \texttt{\bf s}-space at the same time. This method is presented in Fig. \ref{fig:linear_model_performance}. The slope that minimizes the inherent errors is $0.71$ and the expected inherent 1D $1\sigma$ velocity error is then $164\kms$. This analysis takes into account a $10\%$ variability in $N(z)$, these are the grey areas in figures \ref{fig:2mrs_nz} and \ref{fig:linear_model_performance}. Note that the more realistic \textsc{sham} model is not limiting the error; The error is limited by the extreme \textsc{s\_mm} model which includes substructure and only the most massive halos per shell. 
\begin{figure}
	\includegraphics[width=231pt]{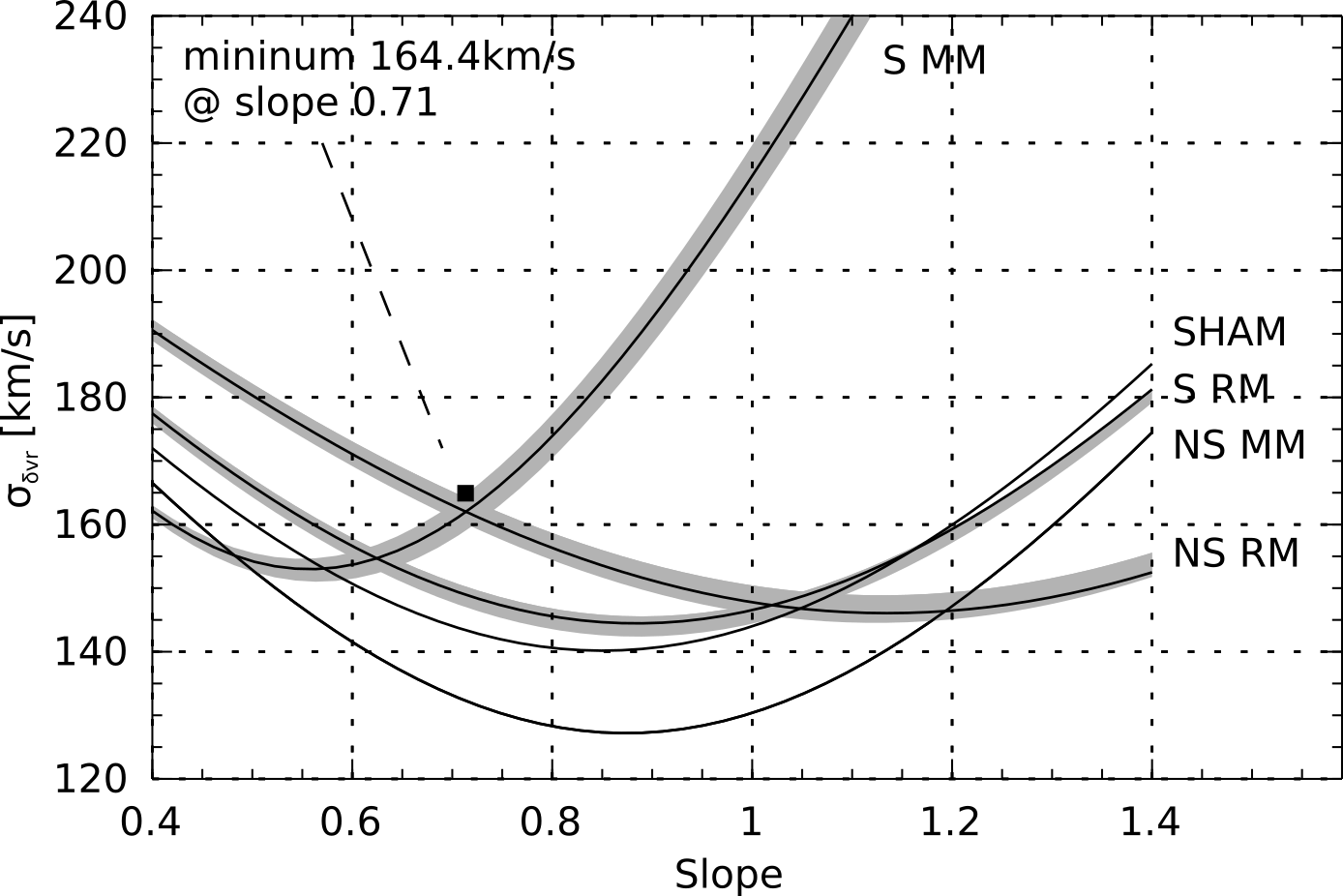} \caption{$\sigma_{{\delta v}r}=\langle(b\vrrec-\vrtru)^2\rangle^{1/2}$ as function of the slope $b$ in the \texttt{\bf s}-space linear reconstruction model. Solid lines correspond to different flux limited catalogs, as indicated in the figure. The scatter around the curves is calculated based on the 30 realizations of each catalog. Grey areas represent $\pm 10$ variations in the 2MRS redshift distribution $N(z)$.} \label{fig:linear_model_performance} 
\end{figure}

\section{Discussion and Conclusions} \label{sec:DISC_AND_CONCLUSIONS} 
\begin{figure*}
	\includegraphics[width=16.7cm]{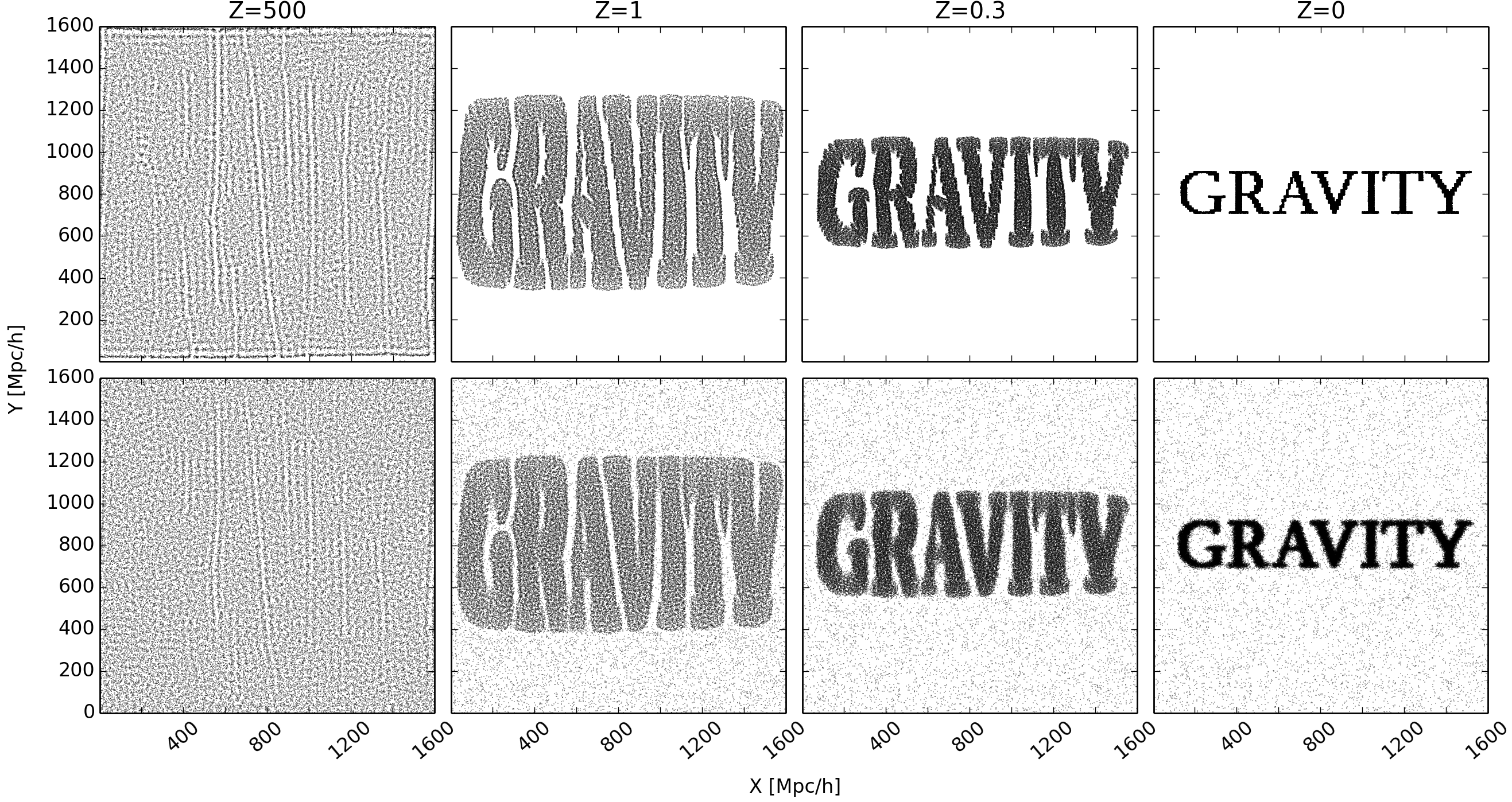} \caption{Two tests of the NoAM method in a periodic box of volume $(1600\mpch)^3$, with $3 \times 10^4$ particles. Panels to the right are target configurations, top and bottom panel with sharp and soft edges respectively, as explained in the text. Left panels correspond to reconstructed ICs at redshift 500, panels in the middle are snapshot configurations resulting from running the ICs forward in time using full N-body dynamics. At redshift $z=0$ the target is visually indistinguishable from the reconstructed configuration: the mean 1D distance between particles to their target is $0.5\mpch$ while mean particle separation is $50\mpch$} \label{fig:GRAVITY} 
\end{figure*}

We have presented a systematic comparison of Lagrangian methods for large scale velocity field reconstruction. We have focused only on the dynamical aspect of the reconstruction. Galaxy surveys are limited in the range of scales that they probe. The boundary of the survey imposes a stringent limit on the largest scales that can be useful for extracting cosmological information. A limit on small scales is dictated by the mean galaxy separation in a given redshift survey\footnote{For example the Euclid survey \citep{2012SPIE.8442E..0TL} BAO optimized survey (NIR) will map 3500 galaxies per square degree with spectroscopic redshift measurements from $z=0.7$ to $2.05$. This gives a mean galaxy separation of about $10\mpch$. The weak lensing optimized survey will record 30 galaxies per squared arcminute with a median redshift of $0.8$, giving a mean galaxy separation of about $3\mpch$, however with a large redshift error $\sigma(z)=0.05(1+z)$.}. A large scale mode is expected to obey linear theory and to remain dynamically decoupled. However, substantial non-linear coupling occurs between modes within the scale of nonlinearities. Therefore, missing information on small scales affects the reconstruction on the observed scales. In addition, observing galaxy redshifts (rather than distances) introduces 
additional severe ambiguities on small scales. 
Observed structures in redshift surveys appear smeared out in the radial direction due to small scale incoherent motions.
Further,   even nearly  laminar flow regions \texttt{\bf r}-space with negligible  small scale random motions    could well  be associated with the presence of multi-values zones\footnote{Matter at different positions in \texttt{\bf r}-space but with the same redshift coordinate}. The spatial extent of both effects is  as large as a few Megaparsecs, even without the inclusion of fingers-of-god from rich galaxy clusters. Thus reconstruction on smaller scales is a very challenging task. We adopt the strategy of assessing the effect of small scales using mock catalogs properly tailored to match the observations. Nonetheless, a variety of methods have been developed which incorporate statistical priors in the reconstructions. These methods typically combine the Zel'dovich and 2LPT reconstructions \citep{1992ApJ...384..448H, 2013MNRAS.435L..78K, 2015JPhA...48v5205L, 2016MNRAS.460.1340M, 2016MNRAS.458..900C}. This class of methods can be useful for  statistical assessments of the effects of missing scales. Our approach is to do that by a calibration with realistic mock catalogs extracted from N-body simulations.

A significant advantage of NoAM is the speed of convergence: in this study we use it on the $1.5\times10^8$ particles of the \itype catalogs in a periodic box of side length $1600\mpch$ with a $3\mpch$ Gaussian gravitational smoothing. Running NoAM with a PM simulation on 32 cores takes 2 days for a convergence threshold of $0.03$. Any distributed N-body code can be used to implement NoAM, and more so - the simulations in NoAM have a gravitational softening equal to the mean particle separation in the survey of interest, which is in the orders of Megaparsecs - much larger than the softening used in structure formation simulations. With 10 nodes as those used here  (320 cores in total)  a similar analysis could be performed for  $10^{10}$ particles in  only a few weeks. NoAM will be able to analyze the richest upcoming galaxy redshift surveys. 

 Two other methods use the action minimization principle to achieve the same goals as NoAM: Path Interchange Zel'dovich Approximation \citep[][hereafter PIZA]{1997MNRAS.285..793C} and the Fast Action Method \citep{Nusser2000, 2002MNRAS.335...53B}. Both methods parametrize particle orbits, PIZA with a Zel'dovich velocity and FAM with a set of $n$ orthogonal basis functions. FAM can be interpreted as a generalization of PIZA, and NoAM can be interpreted as a further generalization of FAM as it can handle parametrized particle orbits but this is not strictly required. Another method for IC reconstruction is named after the Monge--–Amp\'ere--–Kantorovich (MAK) equation \citep{2003A&A...406..393M, 2002Natur.417..260F, 2008MNRAS.383.1292L}. MAK assumes that particle positions at redshift zero are well approximated as a gradient of a convex potential of the initial particle positions $\vec{x}=\vec{\nabla_q}\phi(\vec{q})$ (convexivity is needed for solution uniqueness, and it requires no multistreaming). Using the mass conservation equation $\rho_0d^3q=\rho(\vec{x})d^3x$ the problem reduces to the following Monge--–Amp\'ere--–Kantorovich equation 
\begin{equation}
	\label{eq:MAK_EQ} {\rm det} \left( \nabla_x, \, \nabla_{x_j} \Theta(\vec{x}) \right) = \rho(\vec{x})/\rho_0 
\end{equation}
where convexity allows $\vec{q}=\vec{\nabla_x} \Theta(\vec{x})$. This equation is equivalent to a minimization problem (Brenier 1987; Benamou \& Brenier 2000), the discretized cost function given by 
\begin{equation}
	\label{eq:MAK_DISC} I=\Sigma_{i=1}^{\rm N} \left| \vec{x}_i-\vec{q}_{j(i)} \right|^2 \;. 
\end{equation}
This class of minimizations are called assignment problems, where  observed particles are, respectively, matched  to initial uniformly distributed  points   
on a  predefined grid or glass configurations. The best known algorithm to solve such matching problems (with a quadratic cost function) have a complexity that scales as $\mathcal{O}(n^{2.5})$ (Burkard \& Derigs 1980, Bertsekas 1998). After initial positions are found, peculiar velocities are estimated by first (or second) order Lagrangian perturbation theory as explained in \S \ref{sec:2LPT_ZELD_APPROX}. The computational complexity of the method renders it unfeasible for catalogs with more than a few times $10^5$ galaxies.

We have considered  a volume large enough to allow for  significant cosmic variance in spheres of $150\mpch$ (the \rtype catalogs which mimic the 2MRS), and DM halos are populated with galaxies using different models. Our results are in general agreement with previous results, however a direct comparison with others is not feasible due to the large number of parameters used in the relevant methods. \cite{2002MNRAS.335...53B} compare FAM, PIZA and linear theory in \texttt{\bf s}-space for a flat $\Lambda CDM$ universe with $\Omega_m=0.3$, $\Omega_\Lambda=0.7$ and a high normalization $\sigma_8=1.13$. Respectively, their cartesian X component $\sigma_v$ for the volume limited catalog (particles, smoothed at $5\mpch$, as shown in their table 4) is $280\pm 57\kms$, $254\pm 35\kms$ and $271\pm 36\kms$. For our particles catalog we have $201\pm 12\kms$, $200\pm 11\kms$ and $196\pm 11\kms$. Their numbers are higher than ours, mainly because of the higher normalization. Also notable is that in both cases there is no significant difference between the different methods used. The slope they find is, respectively, $0.26\pm 0.04$, $0.81\pm 0.05$ and $0.98\pm 0.06$, whereas in our analysis all methods give a slope not significantly different from $0.75$. There could be several reasons for the low linear slope found by \cite{2002MNRAS.335...53B}: 
\begin{enumerate*}
	\item the gravitational softening of $0.5\mpch$ vs our $3\mpch$ 
	\item no use of a consistent linear theory velocity in \texttt{\bf s}-space 
	\item the radius of the sphere of $40\mpch$ vs our $150\mpch$. 
\end{enumerate*}

A question that is often raised is wether any clumpy distribution of particles can be obtained by gravitational evolution of homogeneous initial conditions. In NoAM, the set of initial particle velocities can be treated as free parameters and the target positions as the data to be matched by the model. The model connecting the parameters to the data is gravitational dynamics. There is a strong covariance between the free parameters, which results from the force softening. By choosing a sufficiently small force softening length it should be possible to tune the initial velocities to match the observed final positions. The following example aims at demonstrating this point. We generate a highly non-uniform ad hoc target (thus more challenging) which is also easily recognizable by eye. The target particle positions are randomly distributed within a slice of thickness $60\mpch$ in a region defined by the word ``GRAVITY", see Fig. \ref{fig:GRAVITY}. We aim at finding uniform initial conditions which using full N-body dynamics evolve to the target distribution. To do that we run NoAM with $30\times10^3$ particles of equal mass in a box of volume $(1600\mpch)^3$. The cosmological parameters used are this same as in $\S$\ref{sec:THE_NOAM_METHOD}, with a gravitational force softening of $3\mpch$. We run the test twice, the second time with a softened target (thus less challenging). The target is softened by uniformly redistributing $30\%$ of the particles to the whole box volume. For the tests we use ${\rm Tol}=0.01$, $z_{\mathrm ini}=500$ and $0.001<f<0.1$ as typical values. Top and bottom rows in Fig. \ref{fig:GRAVITY} correspond to the targets with sharp and softened edges, respectively. Left panels show the reconstructed ICs which are then run with full N-body dynamics to redshift zero. Snapshots of particle positions are given in the panels to the right, the redshift is noted for each column. In the rightmost panels the particle configuration is indistinguishable (by eye) from the target.

\section*{Acknowledgments} This research was supported by the I-CORE Program of the Planning and Budgeting Committee, THE ISRAEL SCIENCE FOUNDATION (grants No. 1829/12 and No. 203/09) and the Asher Space Research Institute. AN is grateful to the Heidelberg Institute of Theoretical Astrophysics where this project has been finalized. 
\bibliographystyle{mnras} 
\bibliography{KN16.bbl}

\bsp

\label{lastpage}

\end{document}